\documentclass[aoas]{imsart}
 
\RequirePackage{amsthm,amsmath,amsfonts,amssymb,relsize,mathtools}
\RequirePackage[authoryear]{natbib}
\RequirePackage[colorlinks,citecolor=blue,urlcolor=blue]{hyperref}
\RequirePackage{graphicx,float,multirow}

\startlocaldefs
\theoremstyle{plain}

\theoremstyle{remark}
\newtheorem{assumption}{Assumption}

\endlocaldefs

\begin{document}

\begin{frontmatter}
\title{Functional-Coefficient Models for Multivariate Time Series in Designed Experiments: with Applications to Brain Signals}
\runtitle{Mixed-Effects Functional-Coefficient Model for Multivariate Time Series}

\begin{aug}
\author{\fnms{Paolo~Victor}~\snm{Redondo}\ead[label=e1]{paolovictor.redondo@kaust.edu.sa}},
\author{\fnms{Rapha\"el}~\snm{Huser}\ead[label=e2]{raphael.huser@kaust.edu.sa}\orcid{0000-0002-1228-2071}}
\and
\author{\fnms{Hernando}~\snm{Ombao}\ead[label=e3]{hernando.ombao@kaust.edu.sa}}

\address{Statistics Program, Computer, Electrical, and Mathematical Science and Engineering (CEMSE) Division, King Abdullah University of Science and Technology (KAUST), Thuwal 23955-6900, Saudi Arabia\printead[presep={,\\ }]{e1,e2,e3}}

\end{aug}

\begin{abstract}
To study the neurophysiological basis of attention deficit hyperactivity disorder (ADHD), clinicians use electroencephalography (EEG) which record neuronal electrical activity on the cortex. Instead of focusing on single-channel spectral power, a novel framework for investigating interactions (dependence) between channels in the entire network is proposed. Although dependence measures such as coherence and partial directed coherence (PDC) are well explored in studying brain connectivity, these measures only capture linear dependence. Moreover, in designed clinical experiments, these dependence measures are observed to vary across subjects even within a homogeneous group. To address these limitations, we propose the mixed-effects functional-coefficient autoregressive (MXFAR) model which captures between-subject variation by incorporating subject-specific random effects. The advantages of the MXFAR model are the following: (i) it captures potential non-linear dependence between channels; (ii) it is nonparametric and hence flexible and robust to model mis-specification; (iii) it can capture differences between groups when they exist; (iv) it accounts for  variation across subjects; (v) the framework easily incorporates well-known inference methods from mixed-effects models; (vi) it can be generalized to accommodate various covariates and factors. Then, we formulate a novel non-linear spectral measure, the \textit{functional} partial directed coherence (\textit{f}PDC), to extract dynamic cross-dependence patterns at different frequency oscillations. Finally, we apply the proposed MXFAR-\textit{f}PDC framework to analyze multichannel EEG signals and report novel findings on altered brain functional networks in ADHD patients.
\end{abstract}


\end{frontmatter}

\section{Introduction}\label{chap:intro}

Attention deficit/hyperactivity disorder (ADHD) is one of the most common cognitive disorders \citep{alchalabi2018focus} that is mostly prevalent among children \citep{chen2019deep}. Individuals with ADHD show symptoms including locomotor hyperactivity (being overactive), impulsive behavior, excitability, emotional immaturity, short attention span (being inattentive), distractibility, and inefficiency at school or work \citep{gualtieri2005adhd}. Persistence of the disorder until adulthood could lead to poor academic performance and problems in social interactions that may become a lingering concern. In some extreme cases, adult ADHD patients suffer from one or more additional psychiatric disorders on top of mood and anxiety disorders, personality disorders and even substance abuse \citep{sobanski2006psychiatric}. Driven by the severity of the impact of this medical condition, the goal in this paper is to develop a new statistical framework that can be used by clinicians to study alterations in brain function that are associated with ADHD. Precisely, we develop a formal methodology under which we can identify differences in brain functional networks between the healthy controls and the ADHD population using electroencephalography (EEG) data.

Characterization of ADHD remains highly debated as the current descriptors of the disorder are too broad---they lack unique attribution compared to other behavioral and learning disorders, and rely heavily on the reports of behavioral symptoms \citep{swartwood2003eeg}. To develop more objective approaches, many pursue the use of psychophysiological measures such as pupillometry, and neuroimaging modalities such as functional magnetic resonance imaging (fMRI) and EEG. For example, \cite{vimalajeewa2022wavelet} and \cite{li2024anopow} employed wavelet-based and frequency domain approaches on pupil diameter time series to investigate pathological differences between ADHD patients and healthy controls. There are also works focusing on the diagnosis of ADHD among subjects through machine learning techniques applied to fMRI \citep{shao2020classification,cao2023machine} and EEG \citep{moghaddari2020diagnose,taghibeyglou2022detection} data.

For this paper, we focus on EEG. The potential utility of EEG in understanding ADHD is vast because it is easy to collect, relatively inexpensive, suitable for naturalistic experiments and has excellent temporal resolution (at the millisecond scale) \citep{lenartowicz2014use}. This allows one to investigate stimulus-induced changes in brain activity as well as the evolution of the underlying brain process across an entire experiment. The standard approach to analyzing EEG data is to decompose their total variance into the power explained by different frequency bands: delta (0.5–4 Hz), theta (4–8 Hz), alpha (8–12 Hz), beta (13–30 Hz), and gamma (30–100 Hz) \citep{ombao2005slex,nunez2016electroencephalography,fiecas2016modeling,guerrero2023conex}. From the very first findings of slow-wave EEG activity in the theta frequency band \citep{jasper1938electroencephalographic}, recent studies explore the difference in spectral power of the EEG signals from individuals with ADHD and normal (without any behavioral and learning disorder) controls. One of the most popular metrics is the theta-to-beta power ratio (TBR). This is associated with ADHD patients having higher theta power (in the fronto-central brain regions) and lower beta power (temporal regions) compared to healthy controls. However, results from previous works have not yet been replicated and, for this reason, have not reached benchmark status in the community. For an in-depth meta-analysis and comparison of recent TBR research works, see \cite{arns2013decade} and \cite{van2020different}.

Although analysis of spectral power offers an objective view to study the neurophysiological basis of mental diseases, its clinical relevance for ADHD applications, to become widely-accepted, is yet to be established. Thus, a complementary approach to spectral analysis of EEG data from a single channel is to examine the cross-dependence between channels in the brain functional network. Coherence and partial directed coherence (PDC) are two of the most considered dependence measures in the literature on EEG data analysis \citep{ombao2022spectral}. Testing for significant differences in the brain connectivity between people with and without cognitive disorder (such as ADHD and dyslexia) based on these measures provides interesting insights (see for example \citealp{murias2007functional,clarke2007coherence,erkucs2017graph,muthuraman2019multimodal}). However, these measures capture only linear dependence, and hence, one of the major limitations of these metrics is that they fail to identify non-linear dependence structure in cross-channel dependence in the brain network. For instance, non-linear interactions may occur when the general state of the brain, which may be summarized by activities in a representative channel, drives the communication pathways between channels in the brain network. Such a phenomenon is also evident in fMRI, e.g., when BOLD signal interactions become more significant due to BOLD activity from a distant interacting region \citep{lahaye2003functional}, and is closely linked with suppression and activation in brain connectivity \citep{anticevic2012role,xu2016activation}. For this reason, it is more realistic, as empirically demonstrated, to assume a dynamic dependence structure driven by some reference signal, and hence, investigating them via a universal linear measure of dependence becomes inappropriate.

Another challenge is that, since these dependence measures are not directly observed, they have to be estimated for each individual. This introduces a number of disadvantages including neglecting shared attributes across subgroups (e.g., similar ADHD-type) and requiring homogeneity among subject in terms of age, gender, etc. A solution to this issue is to consider a mixed-effects framework as inspired by the works of \cite{diggle1997spectral}, \cite{freyermuth2010tree}, \cite{krafty2011functional} and \cite{li2024anopow}, however developed for the analysis of spectral power of stationary and nonstationary univariate time series. Hence, our goal is to develop a statistical method that can capture potentially non-linear interactions between the channels common within a group and address heterogeneity across subjects in the population.

In this paper, we develop a class of models which simultaneously accounts for (i) the non-linear interactions between channels in a brain network through a functional-coefficient autoregressive (FAR) model and (ii) the variability in brain network across subjects. By considering a reference signal that drives the dynamic dependence structure, the FAR formulation naturally overcomes the limitations of existing linear modeling approaches, e.g., the classical vector autoregressive (VAR) model. To address the well-known problem of substantial variation between subjects, we propose an innovative mixed-effects modeling approach, which incorporates subject-specific random effects in a local linear regression paradigm. This enables for aggregating information that is common to all subjects and hence, deriving a collective summary of the temporal dependencies shared through the entire dataset. An additional advantage of our method is the flexible configuration of fixed effects which allows for rigorous comparison between functional networks of two or more populations. Then, we introduce the new concept of \textit{functional} partial directed coherence (\textit{f}PDC), which is directly derived from the autoregressive coefficient functions of the FAR model. As a non-linear generalization of the original PDC calculated from VAR coefficients, the \textit{f}PDC measure provides a more informative and easy-to-interpret non-linear measure of directional dependence in the frequency domain. Combining our new class of non-linear mixed-effects models with our new spectral measure yields a better framework for studying dynamic interactions in complex systems such as functional brain networks. More importantly, this new framework addresses one of the primary objectives of this paper, i.e., it enables for comparing functional dependence between patient groups while taking into account variation from patients across populations.

The remainder of this paper is organized as follows. Section~\ref{chap:data} provides the description of the dataset that motivated the development of our work. In Section~\ref{chap:mxfar}, we detail the novel elements of the postulated mixed-effects functional-coefficient autoregressive (MXFAR) model and the proposed flexible non-parametric estimation method. Formulation of the new \textit{f}PDC measure and a demonstration of its advantages over linear measures are presented in Section~\ref{chap:fpdc}. To investigate non-linear cross-dependence between channels in an EEG network, we apply the proposed MXFAR-\textit{f}PDC framework in Section~\ref{chap:analysis} and report novel findings and interesting results that identify alterations in brain connectivity related to ADHD. Inference using the MXFAR-\textit{f}PDC framework is implemented in our \texttt{R} package called \texttt{mxfar}, which is made available on the Github repository \url{https://github.com/ptredondo/mxfar}. Lastly, Section~\ref{chap:conclusion} concludes with a summary of the contributions of our proposed methodology and future extensions of our work.

\section{Scientific Problem, Dataset and Pre-processing}\label{chap:data}

\begin{figure}
	\centerline{
		\includegraphics[width=0.7\textwidth]{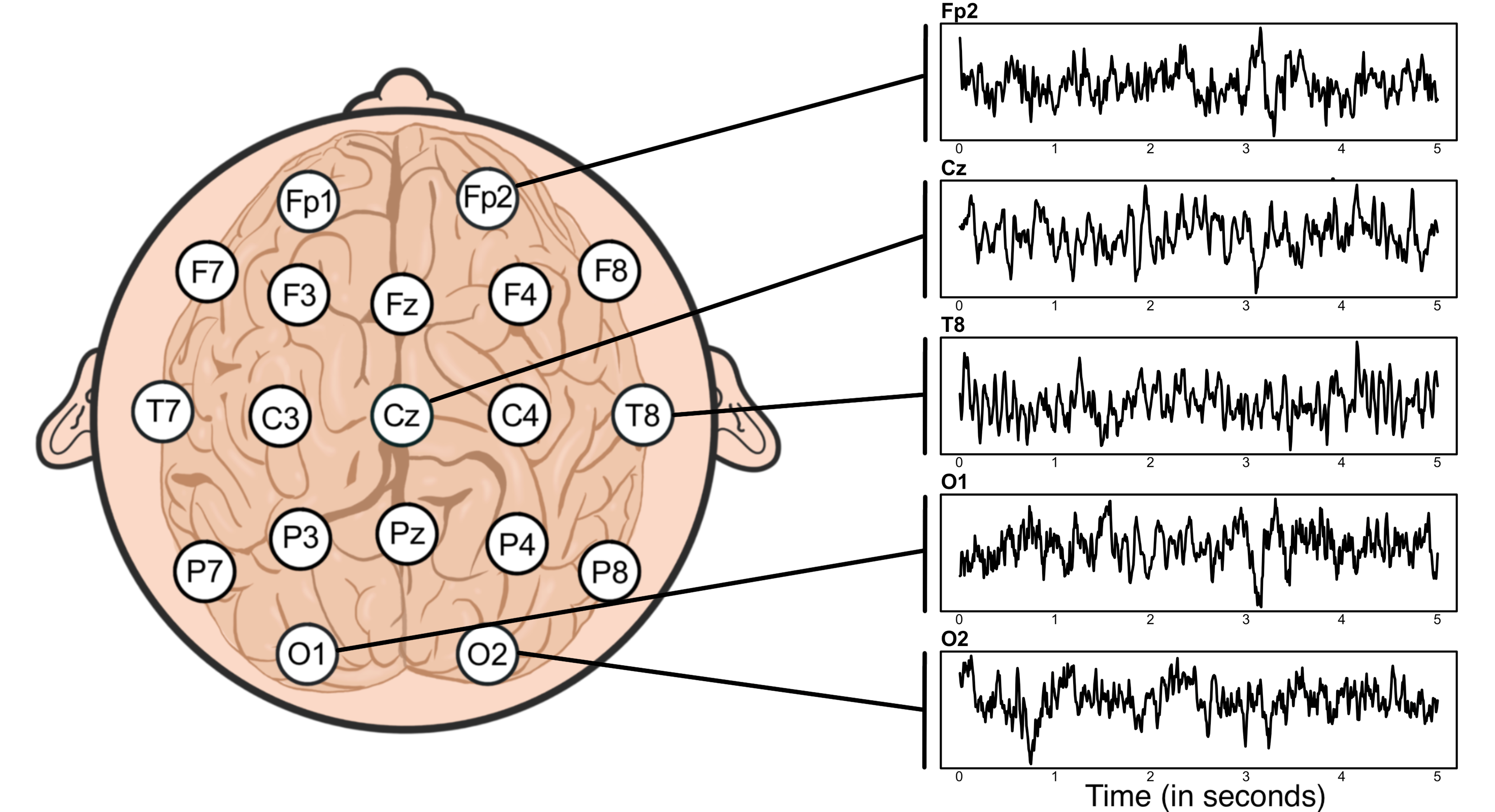} \includegraphics[width=0.3\textwidth]{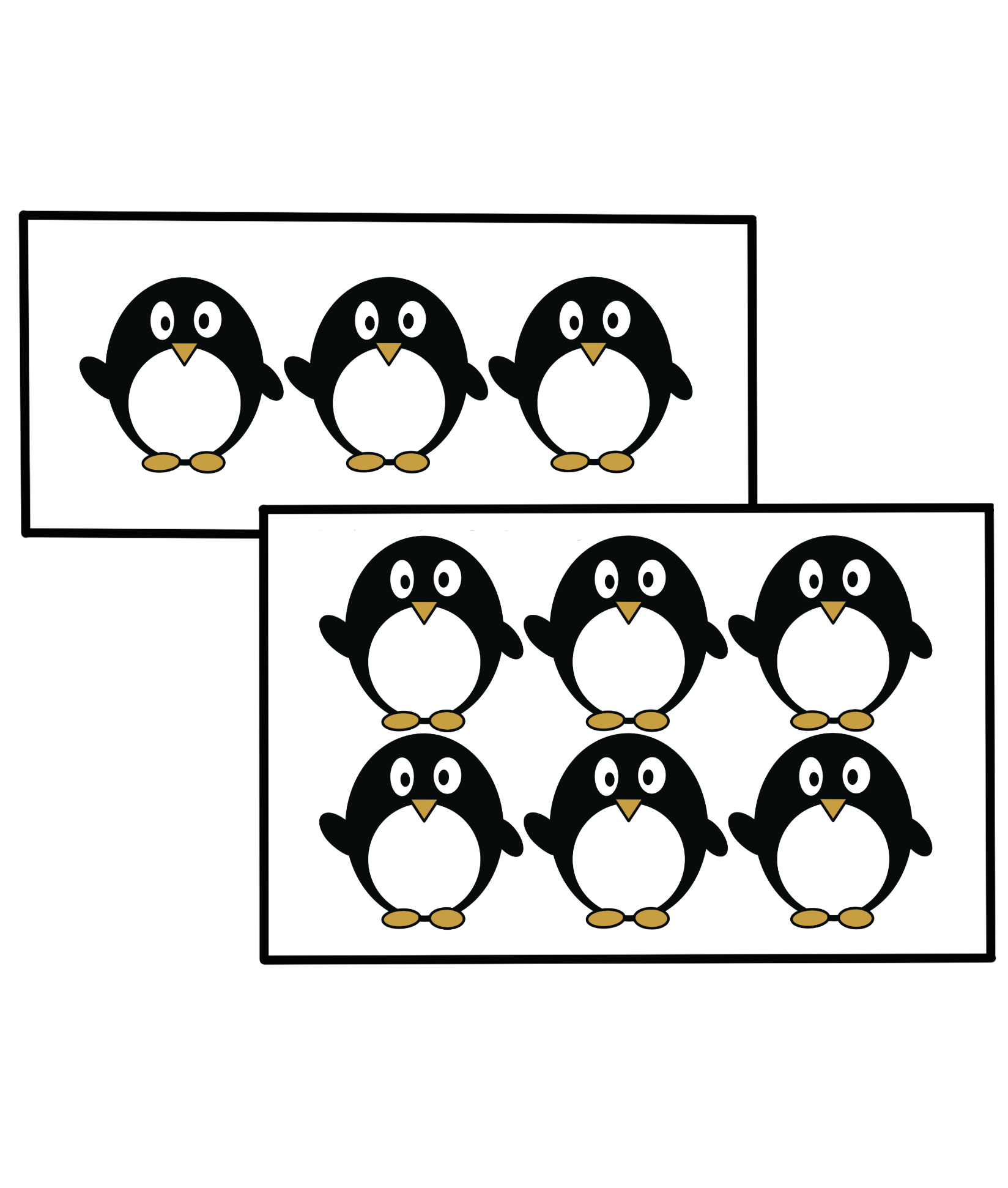} \medskip
	}
	\caption{Standard 19-channel 10--20 EEG scalp topography (left), a 5-second time window (middle) of standardized EEG signals from several channels of a subject with ADHD and an example picture (right) shown to children participants during the experiment.}
	\label{fig:eegsamp}
\end{figure}

The EEG dataset we analyze (reported in \citealp{rzfh-zn36-20}) contains scalp EEG signals sampled at 128Hz from 19 channels in a cognitive experiment with 104 participants: 51 children diagnosed with ADHD and of a control group of 53 children with no registered psychiatric disorder. Note that the original data have 120 subjects but some EEG channels had to be discarded because of severe artifacts and noise that could not be corrected. Pre-processing was conducted via the PREP pipeline of \cite{bigdely2015prep}, which includes artifact removal and bandpass filtering between $0.5$ to $128$ Hertz, to increase the quality of the recordings. Furthermore, the pre-processed EEG signals were standardized to a zero-mean unit variance series to transform all observations to a unified scale. Figure~\ref{fig:eegsamp} shows the standard 10--20 EEG scalp topography and a 5-second time window of standardized EEG signals from selected channels of a subject with ADHD.

The visual-cognitive task in the experiment was to count the number of characters after being shown pictures of cartoon characters (see Figure~\ref{fig:eegsamp}). Seven EEG channels of interest were selected, namely, Fp1 (left pre-frontal), Fp2 (right pre-frontal), O1 (left occipital), O2 (right occipital), T7 (left temporal), T8 (right temporal) and Cz (central), to ``represent'' brain regions that are most likely engaged during the visual task. The frontal region is linked with concentration, focus and problem solving, the occipital region for visual processing, and the temporal region for speech and memory \citep{bjorge2017identification}. On the other hand, the central region is assumed to have a general function since it captures activity over a broad region of the cortex. Here, our objective is to investigate the non-linear cross-dependence patterns in the brain network defined by these selected channels.

In this paper, we use the proposed MXFAR-\textit{f}PDC framework (presented in Sections~\ref{chap:mxfar}~and~\ref{chap:fpdc}) to investigate alterations in brain connectivity among children diagnosed with ADHD, in comparison to children in a healthy control group. There are many challenges to analyzing EEG signals in a designed experiment. First, EEG data are realizations of a highly complex underlying brain functional process. The problem is that most standard time series models are unable to capture non-linear cross-channel interactions in a functional brain network. Second, there is often a significant variation in the brain functional networks across subjects in a population. Thus, to conduct statistical inference on brain network connectivity in these populations that is robust to the subject-specific variations, it is essential to utilize information from multiple subjects and derive a ``common'' functional network that captures non-linear dependencies. A usual strategy is to perform independent analysis per subject and aggregate the subject-specific estimates using some summary measure (e.g., sample mean). However, this does not fully account for the uncertainty of the subject-specific estimates, which assumes equal contribution from each estimates. Hence, it becomes prone to failure as it does not address presence of unusual subjects, even more so in the context of non-linear multivariate time series analysis.

Our novel MXFAR model (see Section~\ref{chap:mxfar}) is specifically tailored to address these issues of non-linear dependence and substantial variation across subjects in a population. By simultaneously estimating the mean and subject-specific functional coefficients through local linear approximation, our proposed method enables for proper utilization of data coming from multiple subjects, and thus, it provides a more natural summarization of the common non-linear dependence structure. Moreover, the straightforward translation of the derived common non-linear dependence structure into the frequency domain, via the \textit{f}PDC measure (see Section~\ref{chap:fpdc}), enables for capturing the complex connectivity patterns at various oscillations that are shared across the subjects. Hence, our MXFAR model and \textit{f}PDC measure are more appropriate tools for analyzing EEG data of individuals from several groups (e.g., ADHD vs. healthy controls).

\section{Mixed-Effects Functional-Coefficient Autoregressive Model}\label{chap:mxfar}

\noindent Consider the EEG signals from $N$ subjects recorded at $k$ different channels and denote them by $\boldsymbol{\underline{Y}}^{(n)}_t = (Y^{(n)}_{1,t},~ Y^{(n)}_{2,t},~ \ldots,~ Y^{(n)}_{k,t})', n = 1, \ldots, N$, where $Y^{(n)}_{j,t}$ represent the EEG recording at channel $j$ and time $t$ from the $n$-th subject. Given the complexity of brain systems, cross-channel interactions are often assumed to exhibit nonlinear dynamics. For instance, connection between the \textit{past} values at channel $g$ and the \textit{future} recordings from the $j$-th channel, i.e., in notation, $Y^{(n)}_g(t) \longrightarrow Y^{(n)}_{j}(t+L)$, for some time lag $L$, may be driven by an external process or other components in the same system. Such a complex dependence dynamics can be captured by the FAR model (see Supplementary Material \citep{redondo2025suppmat} for a concise summary or the original works of \citealp{chen1993functional} and \citealp{harvill2006functional} for the in-depth formulation in the univariate and multivariate settings, respectively). 

Our goal, however, is to capture these nonlinear functional structure in brain networks while accounting for subject-specific variations. That is, we want to utilize information coming from multiple subjects and describe channel interactions that are shared across all subjects. Our solution is to generalize the FAR model to the multi-subject mixed-effects framework. Precisely, we propose a new model which is the mixed-effects functional-coefficient autoregressive (MXFAR) model.
\subsection{Model Definition}\label{chap:MultModel}
A $k$-dimensional time series $\boldsymbol{\underline{Y}}^{(n)}_t$ follows a vector MXFAR process of order $p \in \mathbb{Z}^+$ if, for all $t = 1, 2, \ldots, T$ and $n = 1, \ldots, N$,
\begin{equation}
\boldsymbol{\underline{Y}}^{(n)}_t = \boldsymbol{\mathit{\underline{f}}}^{(n)}(\boldsymbol{\underline{U}}^{(n)}_t)\boldsymbol{\underline{X}}^{(n)}_t + \boldsymbol{\underline{\varepsilon}}^{(n)}_t,~~~~\text{i.e.,}
\label{eq:MultMXFAR}
\end{equation}
\begin{equation*}
\begin{pmatrix}
Y^{(n)}_{1,t} \\
Y^{(n)}_{2,t} \\
\vdots \\
Y^{(n)}_{k,t}
\end{pmatrix} = \begin{pmatrix}
\boldsymbol{\mathit{\underline{f}}}^{(n)}_1(\boldsymbol{\underline{U}}^{(n)}_t) \\
\boldsymbol{\mathit{\underline{f}}}^{(n)}_2(\boldsymbol{\underline{U}}^{(n)}_t) \\
\vdots \\
\boldsymbol{\mathit{\underline{f}}}^{(n)}_k(\boldsymbol{\underline{U}}^{(n)}_t)
\end{pmatrix} \boldsymbol{\underline{X}}^{(n)}_t + \begin{pmatrix}
\varepsilon^{(n)}_{1,t} \\
\varepsilon^{(n)}_{2,t} \\
\vdots \\
\varepsilon^{(n)}_{k,t}
\end{pmatrix},
\end{equation*}
\noindent where $\boldsymbol{\underline{X}}^{(n)}_t = \Big({\boldsymbol{\underline{Y}}^{(n)}_{t-1}}',\ldots,{\boldsymbol{\underline{Y}}^{(n)}_{t-p}}'\Big)'$ with $\boldsymbol{\underline{Y}}_{t-\ell} = (Y_{1,t-\ell},\ldots,Y_{k,t-\ell})', \text{for~} \ell = 1, 2, \ldots, p$, $\boldsymbol{\mathit{\underline{f}}}^{(n)}_j (\boldsymbol{\underline{U}}^{(n)}_t) = \left[f^{(n)}_{j,1:1}(\boldsymbol{\underline{U}}^{(n)}_t),~ \ldots,~ f^{(n)}_{j,k:1}(\boldsymbol{\underline{U}}^{(n)}_t),~ \ldots,~ f^{(n)}_{j,1:p}(\boldsymbol{\underline{U}}^{(n)}_t),~ \ldots,~ f^{(n)}_{j,k:p}(\boldsymbol{\underline{U}}^{(n)}_t)\right]'$, for $j = 1, 2, \ldots, k$, ~$\boldsymbol{\underline{\varepsilon}}^{(n)}_t$ is a zero-mean white noise process with variance-covariance matrix ${\sigma^2_\varepsilon}^{(n)} \boldsymbol{\underline{I}}_k$ with ${\sigma^2_\varepsilon}^{(n)} > 0$, and $\boldsymbol{\underline{\varepsilon}}^{(n)}_t$ is independent of $\boldsymbol{\underline{U}}^{(n)}_s$ for all $s < t$. Here, $\boldsymbol{\underline{U}}^{(n)}_{t}$ may be some vector of exogenous variables, i.e., $\boldsymbol{\underline{U}}^{(n)}_{t} = (U^{(n)}_{1,t},U^{(n)}_{2,t},\ldots, U^{(n)}_{c,t})'$, or an endogenous vector of lagged values of $Y_{j,t}$, e.g., $\boldsymbol{\underline{U}}^{(n)}_{t} = (Y^{(n)}_{t - d_1},Y^{(n)}_{t - d_2},\ldots, Y^{(n)}_{t - d_c})'$, where $j \in \{1,\ldots,k\}$, $d_l > 0, l = 1,2,\ldots,c$. The only condition for $\boldsymbol{\underline{U}}^{(n)}_{t}$ is that its components must be observed and hence, do not rely on any future outcomes. For EEG analysis, we refer to $\boldsymbol{\underline{U}}^{(n)}_{t}$ as the ``reference" signals. Moreover, the functional coefficients $f^{(n)}_{j,g:\ell}(\boldsymbol{\underline{U}}^{(n)}_{t}), j,g = 1,2,\ldots,k,$ $\ell = 1,2,\ldots,p$, which can take on various parametric or nonparametric forms, are measurable functions from $\mathbb{R}^c$ to $\mathbb{R}$, that map a $c$-dimensional vector into the real line. That is, the dependency of $Y_{j,t}$ on the past measurements $Y_{g,t-\ell}$ are functions of the observed reference signals. In neuroscience, the reference signals that drives the brain dynamics may include some deterministic sequence (e.g., convolution of a sequence of stimulus presentation) \citep{rebesco2011stimulus}, recordings from an EEG channel (or a summary from a group of channels) \citep{guerrero2023conex}, or possibly an extraneous physiological signal \citep{di2015dynamic}. For practical purposes and to avoid the \textit{curse-of-dimensionality}, we consider only a single reference signal that governs the nonlinear dependence structure, i.e., we let $\boldsymbol{\underline{U}}^{(n)}_{t} = U^{(n)}_{t}$. Also, we impose the following regularity conditions for the MXFAR model.
\begin{assumption}
    At a given value of the reference signal $U^{(n)}_t = u$, $f^{(n)}_{j,g:\ell}(u)$ is fixed and thus, $\boldsymbol{\underline{Y}}^{(n)}_t$ behaves like a \textit{second-order stationary} vector autoregressive process.
    \label{assump:A1}
\end{assumption}
\begin{assumption}
    For all $j,g = 1, \ldots, k$, $\ell = 1, \ldots, p$ and $n = 1, \ldots, N$, $f^{(n)}_{j,g:\ell}(U^{(n)}_{t})$ is a \textit{smooth function} (i.e., with at least one continuous derivative) of $U^{(n)}_{t}$ where $U^{(n)}_t \in \mathcal{D}_u$ across all $n$, for some support $\mathcal{D}_u$.
    \label{assump:A2}
\end{assumption}
\begin{assumption}
    For $n = 1, \ldots, N$, the functions $f^{(n)}_{j,g:\ell}(\cdot)$ \textit{share the same functional form}, but are governed by some subject-specific parameter vector $\boldsymbol{\underline{\varphi}}^{(n)}_{j,g:\ell}$, i.e.,
    \begin{equation*}
        f^{(n)}_{j,g:\ell}(U^{(n)}_t) = f_{j,g:\ell}(U^{(n)}_t \mid \boldsymbol{\underline{\varphi}}^{(n)}_{j,g:\ell}),
    \end{equation*}
    where $f_{j,g:\ell}(\cdot \mid \cdot)$ is the $\{j,g:\ell\}$-th common coefficient function.
    \label{assump:A3}
\end{assumption}
Assumption~\ref{assump:A1} implies that, locally at $u$, the time series is \textit{well-behaved}, i.e., it has a finite variance, which enables for defining our new frequency-specific nonlinear dependence measure in Section~\ref{chap:fpdc}. In Assumption~\ref{assump:A2}, we require the smoothness of the coefficient functions as we employ a local linear approximation scheme to estimate the MXFAR model (as discussed in Section~\ref{chap:MultEst}), while assuming the same support for $U^{(n)}_t$ across all $n$ allows for formulating a shared dependence structure across the subjects. The former is commonly assumed while the latter is not too difficult to satisfy in practice. For example, considering standardized recordings from the same EEG channel for all subjects as reference signals makes sure Assumption~\ref{assump:A2} is satisfied. Now, allowing $f^{(n)}_{j,g:\ell}(\cdot)$ to take different parametric forms for each $n$, especially when $N$ is large, introduces several issues including (i) the model being too complex to be estimated as it may suffer from overparameterization, and (ii) defining the shared component across subjects becomes vague. Thus, we impose Assumption~\ref{assump:A3}. This is one of the advantages of the proposed MXFAR model. It is straightforward to extract the shared dependence structure between the subjects. Let $\boldsymbol{\underline{\varphi}}^{(\mu)}_{j,g:\ell}$ be the mean of  $\boldsymbol{\underline{\varphi}}^{(n)}_{j,g:\ell}$ across all $n$. Then, we define the $\{j,g:\ell\}$-th mean functional coefficient, denoted by $f^{(\mu)}_{j,g:\ell}(\cdot)$, as the $\{j,g:\ell\}$-th common coefficient function governed by the mean parameter vector $\boldsymbol{\underline{\varphi}}^{(\mu)}_{j,g:\ell}$, i.e., $f^{(\mu)}_{j,g:\ell}(\cdot) = f_{j,g:\ell}(\cdot \mid \boldsymbol{\underline{\varphi}}^{(\mu)}_{j,g:\ell})$. Moreover, it is easy to formulate the MXFAR model to accommodate for multiple groups in the data. For instance, if two groups of subjects are expected to have different dependence structures, mean functional coefficients for each group may be derived. This can be integrated into the model by considering $f^{(\mu_1)}_{j,g:\ell}(\cdot) = f^{1}_{j,g:\ell}(\cdot \mid \boldsymbol{\underline{\varphi}}^{(\mu_1)}_{j,g:\ell})$ and $f^{(\mu_2)}_{j,g:\ell}(\cdot) = f^{2}_{j,g:\ell}(\cdot \mid \boldsymbol{\underline{\varphi}}^{(\mu_2)}_{j,g:\ell})$, where $f^{1}_{j,g:\ell}(\cdot \mid \cdot)$ and $f^{2}_{j,g:\ell}(\cdot \mid \cdot)$ are the common coefficient functions, and $\boldsymbol{\underline{\varphi}}^{(\mu_1)}_{j,g:\ell}$ and $\boldsymbol{\underline{\varphi}}^{(\mu_2)}_{j,g:\ell}$ are the mean parameter vectors for the first and second group, respectively. In the single subject case, i.e., $N = 1$, the MXFAR($p$) model is equivalent to the univariate FAR($p$) model of \cite{chen1993functional} when $k = 1$, while it translates to the vector FAR($p$) model of \cite{harvill2006functional} when $k > 1$, and hence, inherits their nice properties.

Another advantage of the MXFAR model is that it naturally incorporates the variation across subjects in a flexible manner. Suppose, for the $n$-th subject, $\boldsymbol{\underline{\varphi}}^{(n)}_{j,g:\ell} = (\varphi_1,\varphi_2,\varphi^{(n)}_3)'$. Here, $\varphi_1$ and $\varphi_2$ are common to all subjects while only $\varphi^{(n)}_3$ is specific to subject $n$. If $f^{(n)}_{j,g:\ell}(U^{(n)}_t) \coloneqq f_{j,g:\ell}(U^{(n)}_t \mid \boldsymbol{\underline{\varphi}}^{(n)}_{j,g:\ell}) = \varphi_1 \exp\big\{-(\varphi_2 + \varphi^{(n)}_3)(U^{(n)}_t)^2\big\}$, i.e., it takes an exponential form, then the subject-specific component has a multiplicative effect. In smooth transition AR models (the smooth generalization of threshold AR models), if $f^{(n)}_{j,g:\ell}(U^{(n)}_t) = \varphi_1 \left[1 + \exp\big\{\varphi_2(U^{(n)}_t -\eta)\big\}\right]^{-1} + \varphi^{(n)}_3$, then the subject-specific component becomes additive. Thus, our model accounts for the possibility that the dynamics can change from one subject to another, which is frequently encountered in brain imaging data.

\subsection{Estimation}\label{chap:MultEst}

Simultaneously estimating the entire MXFAR model requires extensive computational resources since inclusion of numerous channels ($k$), large number of subjects ($N$) or time points ($T$) leads to inverting extremely large matrices. To overcome this limitation, our approach is to estimate each component separately across channels. For the $j$-th channel in Equation~(\ref{eq:MultMXFAR}), the model is specified to be
\begin{equation}
Y^{(n)}_{j,t} = \boldsymbol{\underline{f}}^{(n)}_j (U^{(n)}_t) \boldsymbol{\underline{X}}^{(n)}_t + \varepsilon^{(n)}_{j,t},
\label{eq:MultMXFARj}
\end{equation}
\noindent with $\boldsymbol{\underline{f}}^{(n)}_j(U^{(n)}_t)$ and $\boldsymbol{\underline{X}}^{(n)}_t$ as defined in Equation~(\ref{eq:MultMXFAR}). For $U^{(n)}_t$ in a small neighborhood around $u$, we can approximate each $f^{(n)}_{j,g:\ell}(U^{(n)}_t)$ locally at $u$ by a linear function, i.e., $f^{(n)}_{j,g:\ell}(U^{(n)}_t) \approx \alpha^{(n)}_{j,g:\ell} + \beta^{(n)}_{j,g:\ell}(U^{(n)}_t - u)$, where $\alpha^{(n)}_{j,g:\ell} = \alpha_{j,g:\ell} + a^{(n)}_{j,g:\ell}$, $\beta^{(n)}_{j,g:\ell} = \beta_{j,g:\ell} + b^{(n)}_{j,g:\ell}$, with the assumption that $\{a^{(n)}_{j,g:\ell},b^{(n)}_{j,g:\ell}\}$ are independent for all $g = 1, \ldots, k$ and $\ell = 1, \ldots,p$, and $a^{(n)}_{j,g:\ell} \sim N(0,\sigma^2_{\alpha_{j,g:\ell}})$ and $b^{(n)}_{j,g:\ell} \sim N(0,\sigma^2_{\beta_{j,g:\ell}})$, for some variances $\sigma^2_{\alpha_{j,g:\ell}} >0 $ and $\sigma^2_{\beta_{j,g:\ell}} > 0$ to be estimated from the data. Thus, we define the local linear estimator as $\widehat{f}^{(n)}_{j,g:\ell}(u) = \widehat{\alpha}_{j,g:\ell} + \widehat{a}^{(n)}_{j,g:\ell}$, where $\big\{\widehat{\alpha}_{j,g:\ell},\widehat{a}^{(n)}_{j,g:\ell},\widehat{\beta}_{j,g:\ell},\widehat{b}^{(n)}_{j,g:\ell}\big\}$ minimizes
$$\sum_{n=1}^{N}\sum_{t=1}^{T} \left[Y^{(n)}_{j,t} - \sum_{\ell=1}^{p}\sum_{g=1}^{k}\{(\alpha_{j,g:\ell} + a^{(n)}_{j,g:\ell}) + (\beta_{j,g:\ell} + b^{(n)}_{j,g:\ell})(U^{(n)}_t - u)\}Y^{(n)}_{g,t-\ell}\right]^2 K_h(U^{(n)}_t-u)$$
\begin{equation}
+ \sum_{n=1}^{N}\sum_{\ell=1}^{p} \sum_{g=1}^{k} \lambda_{\alpha_{j,g:\ell}} (a^{(n)}_{j,g:\ell})^2 + \sum_{n=1}^{N}\sum_{j=1}^{p}\sum_{g=1}^{k} \lambda_{\beta_{j,g:\ell}} (b^{(n)}_{j,g:\ell})^2,
\label{eq:LocalObjFuncME}
\end{equation}
where $K_h(\cdot) = h^{-1}K(\cdot/h)$, with specified kernel function $K$, and fixed bandwidth $h>0$, for some regularization parameters $\lambda_{\alpha_{j,g:\ell}},\lambda_{\beta_{j,g:\ell}} > 0$. In matrix form, (\ref{eq:LocalObjFuncME}) can be expressed as
\begin{equation}
\left[\boldsymbol{\underline{Y}}_j - \left(\boldsymbol{\underline{X}}\boldsymbol{\underline{\theta}}_j + \boldsymbol{\underline{Z}}\boldsymbol{\underline{\gamma}}\right)\right]'\boldsymbol{\underline{W}}\left[\boldsymbol{\underline{Y}}_j - \left(\boldsymbol{\underline{X}}\boldsymbol{\underline{\theta}}_j + \boldsymbol{\underline{Z}}\boldsymbol{\underline{\gamma}}\right)\right] + \boldsymbol{\underline{\gamma}}_j'\boldsymbol{\underline{G}}_{j}^{-1}\boldsymbol{\underline{\gamma}}_j
\label{eq:LocalObjFuncMEMatrix}
\end{equation}
where $\boldsymbol{\underline{Y}}_j = (\boldsymbol{\underline{Y}}_j^{(1)'},\ldots, \boldsymbol{\underline{Y}}_j^{(N)'})'$ with  $\boldsymbol{\underline{Y}}_j^{(n)} = (Y^{(n)}_{j,1},\ldots, Y^{(n)}_{j,T} )'$, $\boldsymbol{\underline{X}} = (\boldsymbol{\underline{\tilde{X}}}^{(1)'},~ \ldots,~ \boldsymbol{\underline{\tilde{X}}}^{(N)'})'$ with $\boldsymbol{\underline{\tilde{X}}}^{(n)}$ as a $T \times 2kp$ matrix such that its $t$-th row is equal to $\left(\boldsymbol{\underline{X}}^{(n)'}_t,~ \boldsymbol{\underline{X}}^{(n)'}_t(U^{(n)}_t - u)\right)$, $\boldsymbol{\underline{Z}}$ is a block diagonal matrix of $\boldsymbol{\underline{\tilde{Z}}}^{(n)} = \boldsymbol{\underline{\tilde{X}}}^{(n)}$, $\boldsymbol{\underline{W}}$ is a block diagonal matrix of subject-specific kernel weights $\boldsymbol{\underline{W}}^{(n)} = \mbox{diag}\{K_h(U^{(n)}_1-u),\ldots,K_h(U^{(n)}_T-u)\}$, $\boldsymbol{\underline{G}}_j^{-1} = \boldsymbol{\underline{\lambda}}_j \otimes \boldsymbol{\underline{I}}_N$ with $\boldsymbol{\underline{\lambda}}_j = \mbox{diag}\{\lambda_{\alpha_{j,1:1}},\lambda_{\beta_{j,1:1}},\ldots,\lambda_{\alpha_{j,k:1}},\lambda_{\beta_{j,k:1}},\ldots,\lambda_{\alpha_{j,1:p}},\lambda_{\beta_{j,1:p}},\ldots,\lambda_{\alpha_{j,k:p}},\lambda_{\beta_{j,k:p}}\}$, $\boldsymbol{\underline{\theta}}_j = ({\boldsymbol{\underline{\alpha}}_j}', {\boldsymbol{\underline{\beta}}_j}')'$, $\boldsymbol{\underline{\alpha}}_j = (\alpha_{j,1:1}, \ldots,\alpha_{j,k:1}, \ldots, \alpha_{j,1:p}, \ldots,\alpha_{j,k:p})'$, $\boldsymbol{\underline{\beta}}_j = (\beta_{j,1:1}, \ldots, \beta_{j,k:1}, \ldots, \beta_{j,1:p}, \\ \ldots, \beta_{j,k:p})'$, and $\boldsymbol{\underline{\gamma}}_j = (\boldsymbol{\underline{\gamma}}_j^{(1)'}, \ldots, \boldsymbol{\underline{\gamma}}_j^{(N)'})'$ with $\boldsymbol{\underline{\gamma}}_j^{(n)} = (\boldsymbol{\underline{a}}_j^{(n)'}, \boldsymbol{\underline{b}}_j^{(n)'})'$ such that~ $\boldsymbol{\underline{a}}_j^{(n)} = (a^{(n)}_{j,1:1}, \ldots, a^{(n)}_{j,k:1}, \ldots, a^{(n)}_{j,1:p}, \ldots, a^{(n)}_{j,k:p})'$ and $\boldsymbol{\underline{b}}_j^{(n)} = (b^{(n)}_{j,1:1}, \ldots, b^{(n)}_{j,k:1}, \ldots, b^{(n)}_{j,1:p}, \ldots, b^{(n)}_{j,k:p})'$. \\

Now, (\ref{eq:LocalObjFuncMEMatrix}) resembles the objective function minimized by the solution of Henderson's mixed model equations \citep{henderson1953estimation}. Therefore, the functional coefficient estimates can be obtained as solutions to 
\begin{equation}
\begin{pmatrix}
\boldsymbol{\underline{X}}'\boldsymbol{\underline{W}}\boldsymbol{\underline{X}} & \boldsymbol{\underline{X}}'\boldsymbol{\underline{W}}\boldsymbol{\underline{Z}} \\
\boldsymbol{\underline{Z}}'\boldsymbol{\underline{W}}\boldsymbol{\underline{X}} & ~~~~\boldsymbol{\underline{Z}}'\boldsymbol{\underline{W}}\boldsymbol{\underline{Z}} + \boldsymbol{\underline{G}}_j^{-1}
\end{pmatrix} \begin{pmatrix}
\boldsymbol{\underline{\theta}}_j \\
\boldsymbol{\underline{\gamma}}_j
\end{pmatrix} = \begin{pmatrix}
\boldsymbol{\underline{X}}'\boldsymbol{\underline{W}}\boldsymbol{\underline{Y}}_j \\
\boldsymbol{\underline{Z}}'\boldsymbol{\underline{W}}\boldsymbol{\underline{Y}}_j
\end{pmatrix}.
\label{eq:HendersonsMMEqcompL}
\end{equation}

Other than the predicted subject-specific functional coefficients $\widehat{f}^{(n)}_{j,g:\ell}(u)$, the estimates for the mean functional coefficients, which is our main interest, are obtained as $\widehat{f}_{j,g:\ell}^{(\mu)}(u) = \widehat{\alpha}_{j,g:\ell}$. Hence, the advantage of our proposed estimation procedure is the extrapolation of the mean behavior of the non-linear dependencies while addressing the variability unique to each subject. Moreover, the proposed estimation method does not require knowledge on the unknown shape or governing parameters of the functional coefficients, which is another benefit of our linear approximation scheme. In addition, our approach can be generalized to account for multiple groups in the data by simply incorporating specific $\alpha_{j,g:\ell}$ and $\beta_{j,g:\ell}$ components for each group. That is, it enables for extracting group mean functional coefficient estimates, say $\widehat{f}_{j,g:\ell}^{(\mu_1)}(u)$ and $\widehat{f}_{j,g:\ell}^{(\mu_2)}(u)$ for two groups, which we exploit in the analysis of EEG data in Section~\ref{chap:analysis}. 

We emphasize that the MXFAR model allows for the coefficient functions $f(\cdot \mid \boldsymbol{\underline{\varphi}}^{(n)}_{j,g:\ell})$, governed by some subject-specific parameter vector $\boldsymbol{\underline{\varphi}}^{(n)}_{j,g:\ell}$, to be arbitrary with possibly additive or multiplicative random effects. However, we approximate them (locally at $u$) by a linear mixed-effects model with additive random effects. Moreover, $\{\hat{\alpha}_{j,g:\ell},\hat{\beta}_{j,g:\ell},\hat{a}^{(n)}_{j,g:\ell},\hat{b}^{(n)}_{j,g:\ell}\}$ are not estimates of the \textit{actual} subject-specific parameters $\boldsymbol{\underline{\varphi}}^{(n)}_{j,g:\ell}$. In fact, it is not our intention to estimate $\boldsymbol{\underline{\varphi}}^{(n)}_{j,g:\ell}$ explicitly. Instead, we want an estimator for the mean and the subject-specific functional coefficients regardless of the actual values of $\boldsymbol{\underline{\varphi}}^{(n)}_{j,g:\ell}$, which we achieve from the proposed inference method. Empirical properties of our procedure are investigated through an extensive simulation study and are reported in the Supplementary Material \citep{redondo2025suppmat}.

There are several considerations regarding the proposed mixed-effects local linear estimation approach. Initially, we assume the error component to be a white noise process. Deviation from this assumption, e.g., presence of autocorrelation in the random errors, may result in a poor fit for the data. Such a problem may be subsumed in the paradigm of functional-coefficient autoregressive moving average (FARMA) models, which to our knowledge, has not yet been proposed in the literature. This is outside the domain of our work, and hence, we leave its development to future research.

Also, since the objective function (\ref{eq:LocalObjFuncME}) suggests estimating the parameters by minimizing the sum of weighted squared residuals while penalizing for the contribution of the random effects $a^{(n)}_{j,g:\ell}$ and $b^{(n)}_{j,g:\ell}$, the role of the tuning parameters $\lambda_{\alpha_{j,g:\ell}}$ and $\lambda_{\beta_{j,g:\ell}}$ becomes crucial. Even though these tuning parameters can be specified individually through some cross-validation method, such a task becomes computationally exhausting given the number of subjects $N$, total data points $T$ and the autoregressive order $p$. Thus, we consider a conservative penalty instead. 

For a specific random effect, say $a^{(n)}_{j,g:\ell}$, we assume its contribution to be inversely proportional to its variance $\sigma^2_{\alpha_{j,g:\ell}}$. Moreover, since the functional coefficients are estimated locally at $u$, the criterion incorporates the weight $K_h(U^{(n)}_t-u)$. That is, we adjust the contribution of the random effect depending on how close the observed value of the reference signal is on the local point $u$. This becomes conservative by considering $\max\limits_{n}\{K_h(U^{(n)}_t-u)\}$. Hence, we propose selecting the tuning parameters as $\lambda_{\alpha_{j,g:\ell}} = \lambda \frac{\sigma^2_{\alpha_{j,g:\ell}}}{\max\limits_{n}\{K_h(U^{(n)}_t-u)\}}$ and $\lambda_{\beta_{j,g:\ell}} = \lambda \frac{\sigma^2_{\beta_{j,g:\ell}}}{\max\limits_{n}\{K_h(U^{(n)}_t-u)\}}$, that is, as a function of a single tuning parameter $\lambda$ scaled by the variation across subjects and maximum distance from the approximation. Selection of $\lambda$ may be done via standard cross-validation techniques where a metric, e.g., the mean squared error (MSE), is minimized. Although this suggestion may result in a sub-optimal fit compared to individually cross-validated tuning parameters, we argue that our approach is more pragmatic, while providing similar inference on the dependence structure, as the latter may take significantly longer time to implement (or may not be even feasible) with large $N$. We illustrate this through a simple example which we discuss in the Supplementary Material \citep{redondo2025suppmat}.

Another consideration is the estimation of the variances of the random effects ($\sigma^2_{\alpha_{j,g:\ell}}$ and $\sigma^2_{\beta_{j,g:\ell}}$, $\ell = 1, \ldots, p$). Our strategy is to employ a two-stage estimation methodology. First, we estimate individual vector FAR($p$) model for each subject using the multivariate local linear regression method proposed by \cite{harvill2006functional} and calculate the variance of the resulting slope and intercept coefficients in the approximating linear functions across all $n$. This provides estimates for $\sigma^2_{\alpha_{j,g:\ell}}$ and $\sigma^2_{\beta_{j,g:\ell}}$ for all $g$ and $\ell$. Then, we obtain the final estimates for the mean and subject-specific functional coefficients using the solution to the Henderson's mixed model equations defined in Equation~(\ref{eq:HendersonsMMEqcompL}). An alternative approach is to use a Bayesian methodology, e.g., the framework implemented in the \texttt{brms} package \citep{RJ-2018-017}, to simultaneously estimate the local linear approximation parameters and the variance of the associated random effects. Such frameworks facilitate deriving posterior distributions for the parameters which allows for quantifying uncertainties of the estimates. However, using Bayesian approaches has its own challenges including optimal selection of prior distributions. Moreover, in designed experiments, its implementation is computationally demanding resulting in a very slow performance in actual data analysis. Thus, we opt for our two-stage approach outlined above that enables for faster inference for the shared dependence structure in the data while accounting for the substantial subject-specific variation, which is the primary goal in this paper.

\subsection{Bandwidth and Reference Signal Selection}\label{chap:MultAPE}

Choosing the appropriate bandwidth is an important aspect of estimating functions via a local linear approximation. A large value of the bandwidth leads to overly smooth estimates of the function (potentially leading to underfitting), while a low bandwidth produces rough estimates (leading to overfitting, i.e., interpolation without summarization). Here, we adapt the modified multi-fold cross-validation method in \cite{cai2000functional} in the multiple subject setting. Consider two positive integers $r$ and $Q$ such that the number of available time points $T > rQ$. We define $Q$ subseries where the $q$-th  subseries is denoted by $\boldsymbol{\underline{\mathcal{Y}}}_q = \{\boldsymbol{\underline{Y}}^{(n)}_{t},n = 1,2,\ldots,N, t = 1,2,\ldots,T-rq\}$ for $q=1,\ldots,Q$. For every $q$, the next steps are to estimate the unknown functional coefficients using a given subseries $\boldsymbol{\underline{\mathcal{Y}}}_q$, and calculate the one-step forecasting error for the succeeding $r$ time points, denoted by $\widehat{Y}^{(n)}_{j,t \mid q}$ for $j = 1, \ldots, k$ and $t = T-rq + 1,\ldots, T-rq + r$. Then, the value of the kernel bandwidth $h$ is selected to be the minimizer of the sum of mean squared prediction errors for all subseries. Formally, define the accumulated prediction error (APE) for the $q$-th subseries as ${\rm{APE}}_q(h) = \sum^{N}_{n=1}\sum^{k}_{j=1}\sum^{T-rq+r}_{t=T-rq+1} (Y^{(n)}_{j,t} - \widehat{Y}^{(n)}_{j,t \mid q})^2$, where $q = 1,2,\ldots,Q$. Then the optimal bandwidth $h_{\text{opt}}$ is the value of $h$ that minimizes the total APE across all $q$ defined as
\begin{equation*}
    {\rm{APE}}(h) = \sum^{Q}_{q=1} {\rm{APE}}_{q}(h).
\label{eq:APE}
\end{equation*}
Here, the one-step ahead predictions $\widehat{Y}^{(n)}_{j,t \mid q}$ are based on functional coefficients computed using the subseries $\boldsymbol{\underline{\mathcal{Y}}}_q$ with bandwidth $h[T/(T-rq)]^{1/5}$, but evaluated at the observed values of the reference signal $\{U^{(n)}_{t}, t = T-rq+1,\ldots,T-rq+r\}$. Such a formulation allows for the contribution of the bandwidth $h$ to vary depending on the number of time points in the subseries. Similar to \cite{harvill2006functional}, we impose having a univariate reference signal across components and use $r = \lfloor0.1T\rfloor$ and $Q=4$ as suggested by \cite{cai2000functional} throughout our application.

This data-driven method can also be used as a guide to simultaneously select the optimal order $p$ for the MXFAR model, the ``best" choice for the reference signal $U^{(n)}_{t}$ (whether a lag value of the series or some exogenous variable), and the delay parameter $d$ for the reference signal if no to little information about the process being modeled exists. For this purpose, it is more appropriate to define APE as function of these tuning parameters, instead of just the bandwidth $h$, i.e., as $\text{APE}(h,p,U_t,d)$. Suppose we consider several candidates for $p \in \{1, \ldots, p_{\text{max}}\}$, for $U_t \in \{Y_{1,t}, \ldots, Y_{k,t}, U_{1,t}, U_{2,t}, \ldots\}$, with $Y_{j,t}$ and $U_{j,t}$ being endogenous and exogenous variables, respectively, and for $d \in \{1, \ldots, d_{\text{max}}\}$. Given a fixed bandwidth $h$, say $h_0$, the candidate $(p_{\text{opt}},U_{\text{opt},t},d_{\text{opt}})$, which is a possible combination of the tuning parameter candidates that minimizes $\text{APE}(h_0,p,U_t,d)$, specifies the ``best" model for the data being analyzed. Then, fine tuning of $h$, via the APE criterion, may be done assuming $p = p_{\text{opt}}, U_t = U_{\text{opt},t}$ and $d = d_{\text{opt}}$. Accuracy of our data-driven approach to select the optimal model is investigated through an extensive simulation study, which we report in detail in the Supplementary Material \citep{redondo2025suppmat}. When computational resources are sufficient, simultaneous selection of all tuning parameters may be carried out, although we expect this to take much longer, especially if the inclusion of $h$ requires searching through a fine grid of possible candidates. Meanwhile, all available prior knowledge (e.g., neurophysiology) should be incorporated in the model to complement the selection suggested by our data-driven approach.

\subsection{Bootstrap-based Non-linearity Test}\label{chap:MultNLT}

Let $\{\boldsymbol{\underline{Y}}^{(n)}_t\}$ be a vector of EEG recordings from $k$ channels that follows an MXFAR($p$) process. If its functional coefficients do not depend on any reference signal, i.e., $\boldsymbol{\mathit{\underline{f}}}^{(n)}(U^{(n)}_t) = \boldsymbol{\underline{\eta}}^{(n)}$, where $\boldsymbol{\underline{\eta}}^{(n)}$ is a matrix of constants independent of $U^{(n)}_t$, then the model reduces to a mixed-effects vector  autoregressive (VAR) model \citep{gorrostieta2012investigating}. This implies linear dependence between channels (conditional on past data, the conditional mean vector is a linear function of past values). To test for the non-linearity of the functional coefficients, consider testing
\begin{equation}
    H_0: \boldsymbol{\mathit{\underline{f}}}^{(n)}(U^{(n)}_t) = \boldsymbol{\underline{\eta}}^{(n)} ~~\text{vs.}~~ H_1: \boldsymbol{\mathit{\underline{f}}}^{(n)}(U^{(n)}_t) \neq \boldsymbol{\underline{\eta}}^{(n)}.
\label{eq:NLTestHyp}
\end{equation}
Under the null hypothesis $H_0$, given an estimator of $\boldsymbol{\underline{\eta}}^{(n)}$, denoted by $\widehat{\boldsymbol{\underline{\eta}}}^{(n)}$, the residual sum of squares (RSS) is defined as
\begin{equation}
    {\rm{RSS}}_0 = \sum^{N}_{n=1}\sum^{T}_{t=1} \left[\boldsymbol{\underline{Y}}^{(n)}_t - \widehat{\boldsymbol{\underline{\eta}}}^{(n)}\boldsymbol{\underline{X}}^{(n)}_t\right]'\left[\boldsymbol{\underline{Y}}^{(n)}_t - \widehat{\boldsymbol{\underline{\eta}}}^{(n)}\boldsymbol{\underline{X}}^{(n)}_t\right],
\label{eq:RSS0}
\end{equation}
while fitting an MXFAR model produces the sum of squared residuals
\begin{equation}
    {\rm{RSS}}_1 = \sum^{N}_{n=1}\sum^{T}_{t=1} \left[\boldsymbol{\underline{Y}}^{(n)}_t - \widehat{\boldsymbol{\mathit{\underline{f}}}}^{(n)}(U^{(n)}_t)\boldsymbol{\underline{X}}^{(n)}_t\right]'\left[\boldsymbol{\underline{Y}}^{(n)}_t - \widehat{\boldsymbol{\mathit{\underline{f}}}}^{(n)}(U^{(n)}_t)\boldsymbol{\underline{X}}^{(n)}_t\right].
\label{eq:RSS1}
\end{equation}

\noindent Then, consider the test statistic $L = {\rm{RSS}}_0/{\rm{RSS}}_1 - 1$.  Clearly, a large value of $L$ provides evidence against the null hypothesis of linear dependence, and hence, rejection of $H_0$ is based on \textit{sufficiently} large values of $L$. To avoid assuming an exact reference distribution for $L$, we develop a new bootstrap procedure in the mixed-effects setting. The outline of the procedure is provided below:
\begin{enumerate}

    \item Estimate an MXFAR($p$) model for the observed data and define the collection of subject-specific centered residuals $\{\boldsymbol{\underline{r}}^{(n)}_t - \boldsymbol{\bar{\underline{r}}}^{(n)}\}$ with $\boldsymbol{\bar{\underline{r}}}^{(n)}$ as the $k$-dimensional mean vector of the model residuals, i.e., $\boldsymbol{\bar{\underline{r}}}^{(n)} = \frac{1}{T-p}\sum^{T}_{t=p+1} \boldsymbol{\underline{r}}^{(n)}_t$ where $\boldsymbol{\underline{r}}^{(n)}_t = \boldsymbol{\underline{Y}}^{(n)}_t - \widehat{\boldsymbol{\mathit{\underline{f}}}}^{(n)}(U^{(n)}_t)\boldsymbol{\underline{X}}^{(n)}_t$ for $t = p+1,\ldots,T$.
    \item Sample with replacement bootstrap residuals from the collection of centered residuals and construct the bootstrap sample
    $$\boldsymbol{\underline{Y}}^{(n),b}_t = \widehat{\boldsymbol{\underline{\eta}}}^{(n)}\boldsymbol{\underline{X}}^{(n)}_t + \boldsymbol{\underline{r}}^{(n),b}_t.$$
    \item Compute the test statistic $L$ by replacing the observed data $\boldsymbol{\underline{Y}}^{(n)}_t$ with the bootstrap sample $\boldsymbol{\underline{Y}}^{(n),b}_t$ in (\ref{eq:RSS0}) and (\ref{eq:RSS1}). Denote it by $L^b$.
    \item Calculate the $p$-value of the test as the relative frequency of the event $\{L^b \geq L\}$ among all bootstrap replicates considered ($b = 1, \ldots, B$).
    \item Given a specified level of significance $\alpha \in (0,1)$, reject the null hypothesis if the $p$-value is less than $\alpha$. Otherwise, do not reject $H_0$.
    
\end{enumerate}

Our proposed bootstrap-based non-linearity test is the direct application of the procedures of \cite{cai2000functional} and \cite{harvill2006functional} in the multiple subjects paradigm. When $N=1$ (i.e., for the single subject case), our test yields the same inference decision (i.e., whether to reject or not reject $H_0$) as the tests of \cite{cai2000functional} and \cite{harvill2006functional} for the univariate and multivariate settings, respectively. Rejecting the null hypothesis indicates presence of non-linear dependence in the time series. In the analysis of EEGs, this suggests significance of the impact of a chosen reference signal in driving the cross-channel interactions. For such cases, a more general dependence measure for brain connectivity, e.g., our proposed MXFAR model, should be explored instead of the existing measures that only capture linear dependence.

\subsection{Computational Considerations, Typical Run Times and Limitations}\label{chap:CompLim}

The task here is to characterize network dependence in a $k$-dimensional vector time series from $N$ subjects through an MXFAR($p$) model. For any given value $u$ in the support of a chosen reference signal, estimation of the corresponding functional coefficients $\boldsymbol{\mathit{\underline{f}}}^{(n)}_j (u), j = 1,2,\ldots,k$, requires manipulation of Henderson's mixed model equations. Specifically, the naive way of calculating the solution involves inverting a $2kp(N+1) \times 2kp(N+1)$ matrix. Obviously, the computational resources needed to accomplish this task grows fast with the dimension of the vector time series $k$, the number of lags $p$ considered and the number of subjects $N$. A more efficient way to calculate the solution is to use the general formula for matrix inversion in block form \citep{petersen2008matrix}. As a result, we can translate the large matrix inversion problem into simply inverting multiple $2kp \times 2kp$ matrices instead. Hence, we transform the computational requirement as a linear function of $N$. However, the computing time is still an exponential function of $k$. With this, we limit our application to the EEG dataset by considering only six channels driven by a single reference channel. The selection of EEG channels is discussed in Section~\ref{chap:data}.

On the other hand, residual computation requires taking the difference between the actual value and the model prediction of the response at specific time $t$. A prior step to this is to estimate and evaluate the functional coefficients given the corresponding value of the reference signal. For example, if the chosen reference signal is $Y_{j,t-d}$, computing model residuals demands estimating the functional coefficients at all observed distinct values of $Y_{j,t-d}$. This is computationally expensive. Thus, we suggest simply segmenting the support of the reference signal into $M$ equal-length intervals and estimating only at discretized representative points. It implies that for an actual value of the reference signal, we find the segment it belongs to and approximate its coefficients using the representative estimate for that interval. Suppose we choose to form $M=50$ non-overlapping intervals, this reduces the number of times we need to estimate the coefficients from $N \times T$ (assuming all values are distinct) to only $M = 50$ times.

\begin{figure}
    \centering
    \includegraphics[width=0.65\linewidth]{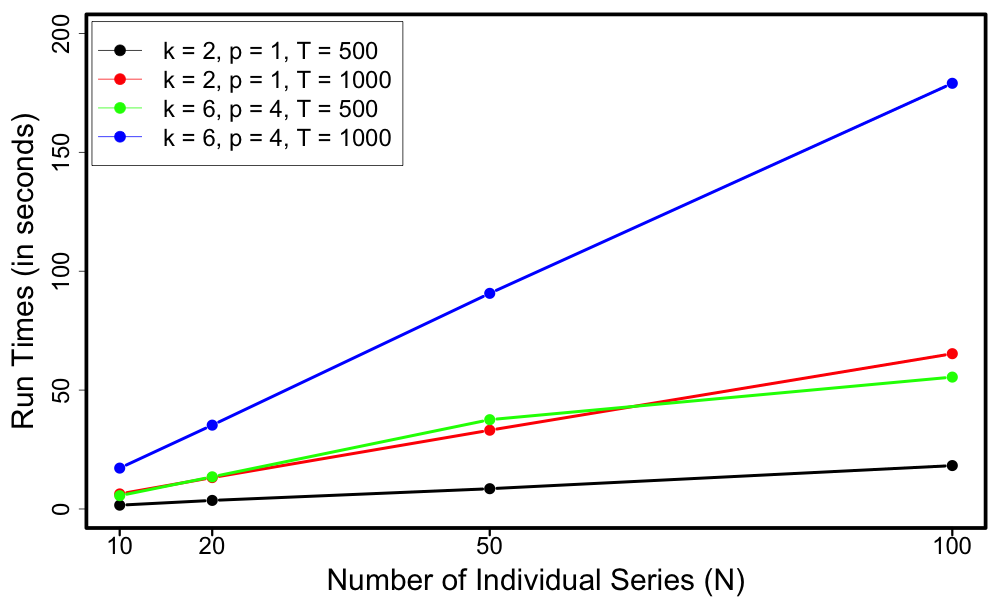}
    \caption{Average run times for estimating a $k$-dimensional vector MXFAR($p$) model over $M = 50$ discretized points of the reference signal.}
    \label{fig:runtimes}
\end{figure}

In Figure~\ref{fig:runtimes}, we show the average run times for estimating the coefficient functions of an MXFAR model using a single central processing unit (CPU) of a regular workstation (i.e., a MacBook Pro equipped with an Apple M1 Pro chip and 16 GB of unified memory). The proposed estimation method provides inference fairly fast for reasonable model and data sizes. However, it slows down for larger model specifications, as expected. We do not report here the run times for the model selection based on the APE metric and the nonparametric test for nonlinearity as these procedures also depend on the number of possible model choices and required bootstrap replicates, respectively. Nonetheless, we emphasize that their respective run times mainly depend on how fast estimation can be done, which we have shown to be decent, and may be sped up through appropriate parallelized computing methods.

\section{Functional Partial Directed Coherence}\label{chap:fpdc}

Another interest is to describe the underlying complex cross-channel interactions at different frequency oscillations that potentially associate to various neurocognitive functions. Thus, the goal here is to develop a non-linear spectral dependence measure which characterizes the strength and direction of the relationship between nodes in the network at a given frequency $\omega \in (0,0.5)$. Since FAR models offer an appealing approach to expressing non-linear dependence via the contribution of a reference signal, our strategy is to formulate the new measure based on its functional coefficients, analogously to the original partial directed coherence (PDC) measure of \cite{baccala2001partial} derived from the coefficients of a linear VAR model.

Formally, we define our proposed frequency-specific measure, which we call the \textit{functional} partial directed coherence (\textit{f}PDC), as follows: Suppose $\{\boldsymbol{\underline{Y}}_t\}$ follows a vector MXFAR($p$) model with $N = 1$, i.e., a vector FAR($p$) process of \cite{harvill2006functional}, and given a value $u$ of the reference signal, denote the functional coefficients at $u$ as $f_{j,g:\ell}(u)$ for $j,g = 1, \ldots, k$ and $\ell = 1, \ldots, p$. Define, for $\omega \in (0,0.5)$,
\begin{equation}
\overline{f}_{j,g}(\omega,u) = 	\begin{cases} 
	1 - \sum_{\ell=1}^{p} f_{j,g:\ell}(u) e^{-i2\pi \omega \ell} & \mbox{if} ~~j = g\\
	- \sum_{\ell=1}^{p} f_{j,g:\ell}(u) e^{-i2\pi \omega \ell} & \mbox{otherwise},\\
	\end{cases}
\label{eq:Fouriercoef}
\end{equation}
\noindent which is the Fourier transform of the functional coefficients. Then, the \textit{f}PDC at frequency $\omega$, given a reference signal value $u$, from channel $g$ to $j$ is calculated as
\begin{equation}
f{\rm{PDC}}_{j,g}(\omega,u) = \frac{\overline{f}_{j,g}(\omega,u)}{\sqrt{\sum_{j' = 1}^{k} \left|\overline{f}_{j',g}(\omega,u)\right|^2}}.
\label{eq:fPDC}
\end{equation}

By translating the functional coefficients $f_{j,g:\ell}(u)$ to $f{\rm{PDC}}_{j,g}(\omega,u)$, we now quantify the intensity and direction of information flow across channel $g$ to $j$ at the specific frequency $\omega$, conditional on the reference signal level $u$. That is, $f{\rm{PDC}}_{j,g}(\omega,u)$ indicates the transfer of information from the $\omega$-oscillations at the previous time in channel $g$ to the future $\omega$-oscillations in channel $j$ that is driven by the reference signal level $u$. Moreover, when the functional coefficients are independent of the reference signal, i.e., $f_{j,g:\ell}(u) = \varphi_{j,g:\ell}$ is constant for all $j, g$ and $\ell$, our \textit{f}PDC measure becomes equivalent to the original PDC, implying $f{\rm{PDC}}_{j,g}(\omega,u) = {\rm{PDC}}_{j,g}(\omega)$ for any $u$. Hence, the \textit{f}PDC can be viewed as a non-linear generalization of the linear PDC for cases when there is a reference signal that drives the non-linear dependence structure. This now provides more insights as the \textit{f}PDC inherits the non-linearity of the FAR model and gains the interpretation and nice properties of the original PDC, e.g., its modulus, $|f{\rm{PDC}}_{\textit{j,g}}(\omega,u)|$, also takes on values in $[0,1]$. For data applications, the \textit{f}PDC measure is calculated by substituting all $f_{j,g:\ell}(u)$ quantities by estimated coefficient functions $\widehat{f}_{j,g:\ell}(u)$.

As existence of PDC requires that the time series follows a second-order stationary VAR process, a similar assumption must hold true for \textit{f}PDC to exist, i.e., stationarity of the vector FAR model. However, general conditions under which a FAR model is stationary remains an open question in spite of almost three decades passing since its first development in the seminal work of \cite{chen1993functional}, though some considerations are provided in \cite{cai2000functional}. When the stationary assumption is violated, the Fourier transform of the functional coefficients in Equation~(\ref{eq:Fouriercoef}) may not exist, which then leads to undefined values for the \textit{f}PDC measure. Nonetheless, when reasonable \textit{f}PDC estimates are obtained from the data (assuming stationarity of the series holds), our proposed metric becomes a better tool for analyzing brain connectivity because it can characterize non-linear interactions in the frequency domain, i.e., beyond linear dependence between oscillatory components of the time series.

\begin{figure}
    \centering
    \includegraphics[width=0.85\textwidth]{figs/FPDC_Example.png}
	\caption{Non-linear spectral dependence between $Y_1$ and $Y_2$ (in both directions) in examples (a) and (b), defined by the \textit{f}PDC (first row) and estimated by the linear PDC (second row), at frequency $\omega \in (0,0.5)$ for different values $u$ of the reference signal, and comparison of \textit{f}PDC and linear PDC (third row) at specific values of $u$.}
	\label{fig:fPDC_Example}
\end{figure}

To highlight advantages of our \textit{f}PDC measure over the linear PDC metric, we consider two motivating examples. Suppose $\{\boldsymbol{\underline{Y}}_t\}$ follows a bivariate FAR($p=1$) process with a single reference signal $\{U_t\}$, i.e.,
\begin{equation*}
\begin{pmatrix}
	Y_{1,t} \\
	Y_{2,t}
	\end{pmatrix} = \begin{pmatrix}
	f_{1,1}(U_t) & ~~~~f_{1,2}(U_t) \\
	f_{2,1}(U_t) & ~~~~f_{2,2}(U_t)
	\end{pmatrix}\begin{pmatrix}
	Y_{1,t-1} \\
	Y_{2,t-1}
	\end{pmatrix} + \begin{pmatrix}
	\varepsilon_{1,t} \\
	\varepsilon_{2,t}
	\end{pmatrix}    
\end{equation*}
with the following functional coefficients:

\begin{itemize}
    \item[(a)] $f_{1,1}(u) = 0.7 - \frac{1.4}{1+e^{15(u-1.5)}}$, ~~~~~$f_{1,2}(u) = 0.6 -  \frac{1.1}{1+e^{4(u-1.5)}}$
    
    \noindent ~~$f_{2,1}(u) = \frac{1.2}{1+e^{5(u-1.5)}} - 0.65$, ~~~~$f_{2,2}(u) = \frac{1.5}{1+e^{15(u-1.5)}} - 0.75$ ~~~where~~~$u \in (0,3);$ \\
    
    \item[(b)] $f_{1,1}(u) = -0.7$, ~~~~~~~~~~~~~~$f_{1,2}(u) = 0.85 e^{-5u^2}$
    
    \noindent ~~$f_{2,1}(u) = -0.6 e^{-4u^2}$, ~~~~~$f_{2,2}(u) = 0.75$ ~~~where~~~$u \in (-1.5,1.5).$
    
\end{itemize}

\noindent The first example (with sigmoidal functional coefficients) represents an abrupt change in the dynamic dependence structure when the reference signal exceeds a certain threshold level. On the other hand, the bivariate exponential autoregressive model relates to the phenomenon of suppression and activation in brain connectivity, e.g., cross-channel interactions appear only when the reference signal exhibits small amplitudes (a possible indication of suppressed brain activity). Such cases are typical behaviors that may be observed from analyzing EEG signals.

Figure~\ref{fig:fPDC_Example} presents the non-linear spectral dependence, demonstrated by the \textit{f}PDC, between the two components of $\boldsymbol{\underline{Y}}_t$ in each example, and the limitations of the original PDC in the presence of a reference signal that drives the dependence structure. In the first example, strong information flow from $Y_2$ to $Y_1$ occurs only during two premises; at low frequencies when small-amplitude reference signals are observed (i.e., when $u < 1.5$), and at high frequencies when the reference signals yields large amplitudes (i.e., when $u \geq 1.5$). The reverse is true for the flow of information from $Y_1$ to $Y_2$. Describing these complex patterns using the linear PDC results in an incomplete characterization of dependence. Specific to this example, the PDC captures only the information flow from $Y_2$ to $Y_1$ at low frequencies and from $Y_1$ to $Y_2$ at high frequencies, but fails to extract interactions specific to other frequency oscillations given large amplitudes of the reference signal. In the second example, the cross-component interactions, i.e., $Y_2 \rightarrow Y_1$ at low frequencies and $Y_1 \rightarrow Y_2$ at high frequencies, occurs only when the magnitude (absolute value) of the reference signal is small. Although the PDC illustrates these interactions at their respective frequencies, it underestimates the strength of information transfer between the two components. In practice, these limitations, which are properly addressed by the \textit{f}PDC measure, may result in missing important cross-channel interactions that critically describes brain connectivity.

Another advantage of the \textit{f}PDC measure is its adaptability to handling data from designed experiments. Precisely, an aggregated non-linear spectral dependence across multiple subjects may be obtained by defining the \textit{f}PDC measure based on the mean functional coefficients $f^{(\mu)}_{j,g:\ell}(u)$ from a fitted vector MXFAR($p$) model. By doing so, between-subject variability is also incorporated in the summarized metric. Thus, combining the MXFAR model and the \textit{f}PDC measure, which we call the MXFAR-\textit{f}PDC framework, enables for extracting the shared non-linear dependence across multiple subjects in the frequency domain. With this, non-linear frequency-specific graphical structures may be constructed; we exploit them in Section~\ref{chap:analysis} to derive the brain networks of children with and without ADHD during a visual task and differentiate the inter-channel connectivity between them.

\section{EEG Analysis: Exploring Non-Linear Dependence in ADHD Data}\label{chap:analysis}

\subsection{Application of the MXFAR-\textit{f}PDC Framework}

From the dataset described in Section~\ref{chap:data}, we explore the evolution of brain connectivity during performance of the visual experiment by employing a sliding window approach. Specifically, we consider 17 time windows where each window has a duration of 5 seconds with a 2.5-second overlap with its adjacent time windows. For each window, we fit a vector MXFAR($p$) model of dimension $k=6$ for the signals collected from the selected EEG channels (namely, Fp1, Fp2, O1, O2, T7 and T8) of 51 children with ADHD and 53 healthy controls.

To determine the best choice for the reference signal, its delay lag $d$ and the model lag order $p$ in the MXFAR model, the extended APE criterion (defined in Section~\ref{chap:MultAPE}) was employed. By examining various combinations of possible reference signal, and values for $d$ and $p$, this step selects the model specifications with the lowest APE value while yielding significant non-linear dependence. Ultimately, the MXFAR process of order $p=4$, with the amplitude of the signals from the central channel Cz (at lag $d=6$) as the reference signal, appeared to be the most appropriate. Moreover, it not only produced the smallest APE across all considered time windows, but channel Cz is also an intuitive choice for the reference signal that drives the non-linear dependence between other channels: because it covers the broadest area at the center-most region of the scalp, Cz serves as the representative for all other channels, especially that the six chosen channels are located at the outermost parts of the neural cortex. Hence, the fitted MXFAR model describes the non-linear connectivity patterns between the pre-frontal, temporal, and occipital channels that are driven by the past activities of other components in the brain network (represented through the central channel Cz).

\begin{figure}
    \centering
    \includegraphics[width=0.6\textwidth]{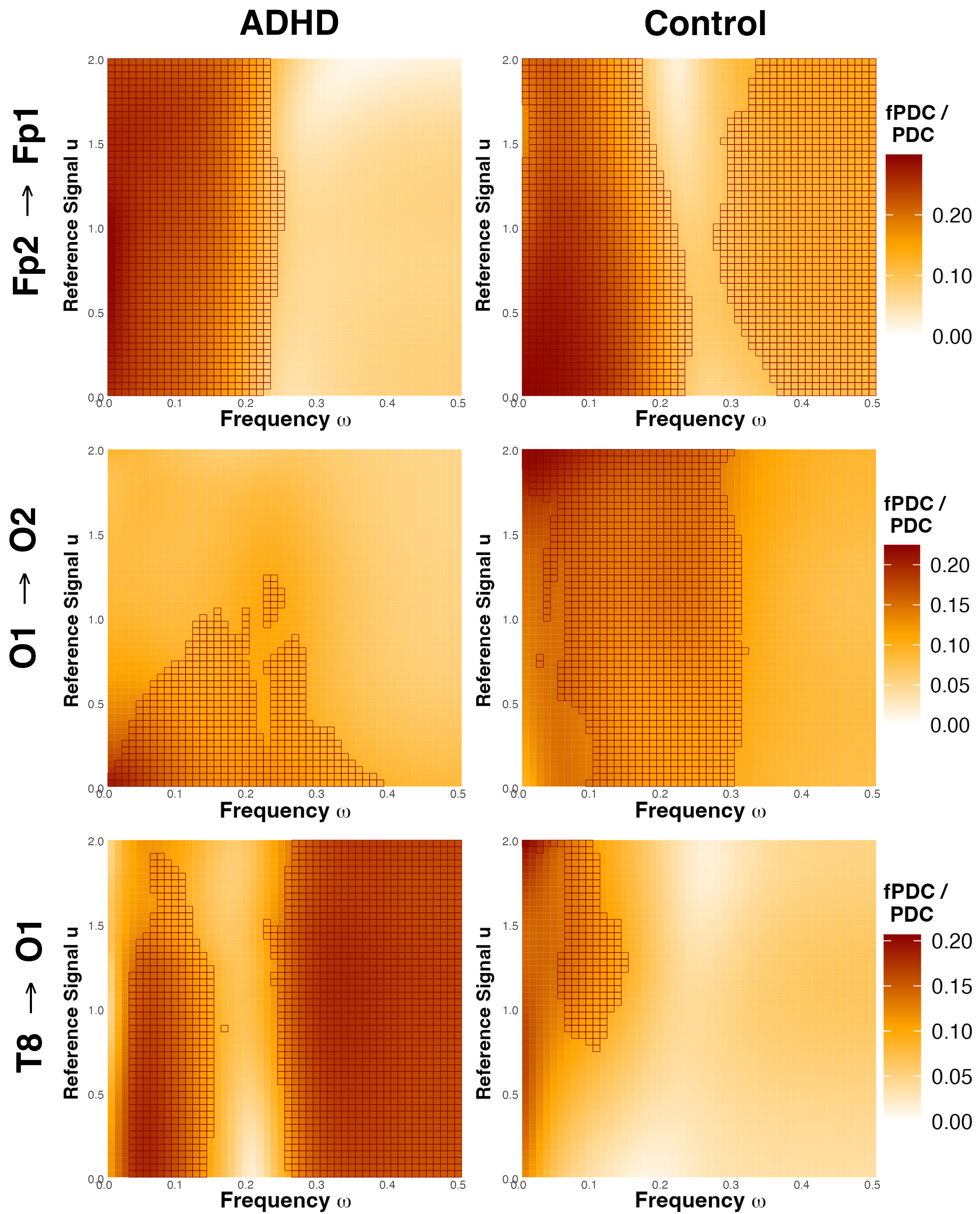}
	\caption{Mean \textit{f}PDC estimates for the ADHD group (left) and control group (right) from several channel pairs (from top to bottom; Fp2~$\rightarrow$~Fp1, O1~$\rightarrow$~O2, T8~$\rightarrow$~O1) for all frequencies $\omega \in (0,0.5)$ given the amplitude $u$ of the reference signal (Cz at lag 6). The shaded regions represent prominent \textit{f}PDC estimates.}
	\label{fig:fPDC_Ex}
\end{figure}

Then, the estimated mean functional coefficients of the MXFAR model for the two groups are transformed into the \textit{f}PDC metric (see Figure~\ref{fig:fPDC_Ex} for example). All remaining group \textit{f}PDC plots for each time window are reported in the Supplementary Material \citep{redondo2025suppmat}. At frequency $\omega~\in~(0,0.5)$ and reference signal value $u$, we say that the directional link from channel $g$ to $j$, denoted by $g \rightarrow j$, is \textit{prominent} if $f{\rm{PDC}}_{j,g}(\omega,u) \geq q_{0.8}(\omega,u)$, where $q_{0.8}(\omega,u)$ is the 0.8-quantile \textit{f}PDC value across all $j$ and $g$ such that $j \neq g$. That is, a directional link becomes prominent if its magnitude is relatively higher compared to other connections in the system.

Now, we divide the analysis of the \textit{f}PDC metric based on different regions of interest for the amplitude of the reference signal and the frequency oscillations. To be precise, we consider the case of small-amplitude reference signals ($u < 1$), i.e., the amplitude is less than one standard deviation away from zero, and large-amplitude reference signals ($u \geq 1$). This enables for visualizing the non-linearity in dependence structure driven by how small or large the past activities are in the other parts of the brain, as represented by the reference signals from channel Cz (at lag $d=6$). On the other hand, we examine two regions in the frequency domain; the low frequency oscillations consisting of the delta, theta and alpha bands, and the high frequency oscillations consisting of the beta and gamma bands. Since the delta, theta and alpha bands are associated with sustained attention and filtering out irrelevant information when performing a mental task \citep{fernandez1995eeg,harmony1996eeg,klimesch1998induced,clayton2015roles,behzadnia2017eeg,van2019functional}, we interpret prominent connections in the low frequencies to be crucial drivers for maintaining focus and attention. Similarly, because beta and gamma oscillations relate to markers of cognitive processing and informational integration \citep{miltner1999coherence,jensen2007human,she2012eeg,gola2013eeg}, we associate directional links in the high frequencies to conscious thinking, i.e., when an individual is voluntarily performing mental actions rather than being in a state of absent-mindedness. Our interests focus on four regions; (i) small-amplitude reference signal at low frequencies, (ii) small-amplitude reference signal at high frequencies, (iii) large-amplitude reference signal at low frequencies, and (iv) large-amplitude reference signal at high frequencies.

\begin{figure}
	\centerline{
		\includegraphics[width=0.8\textwidth]{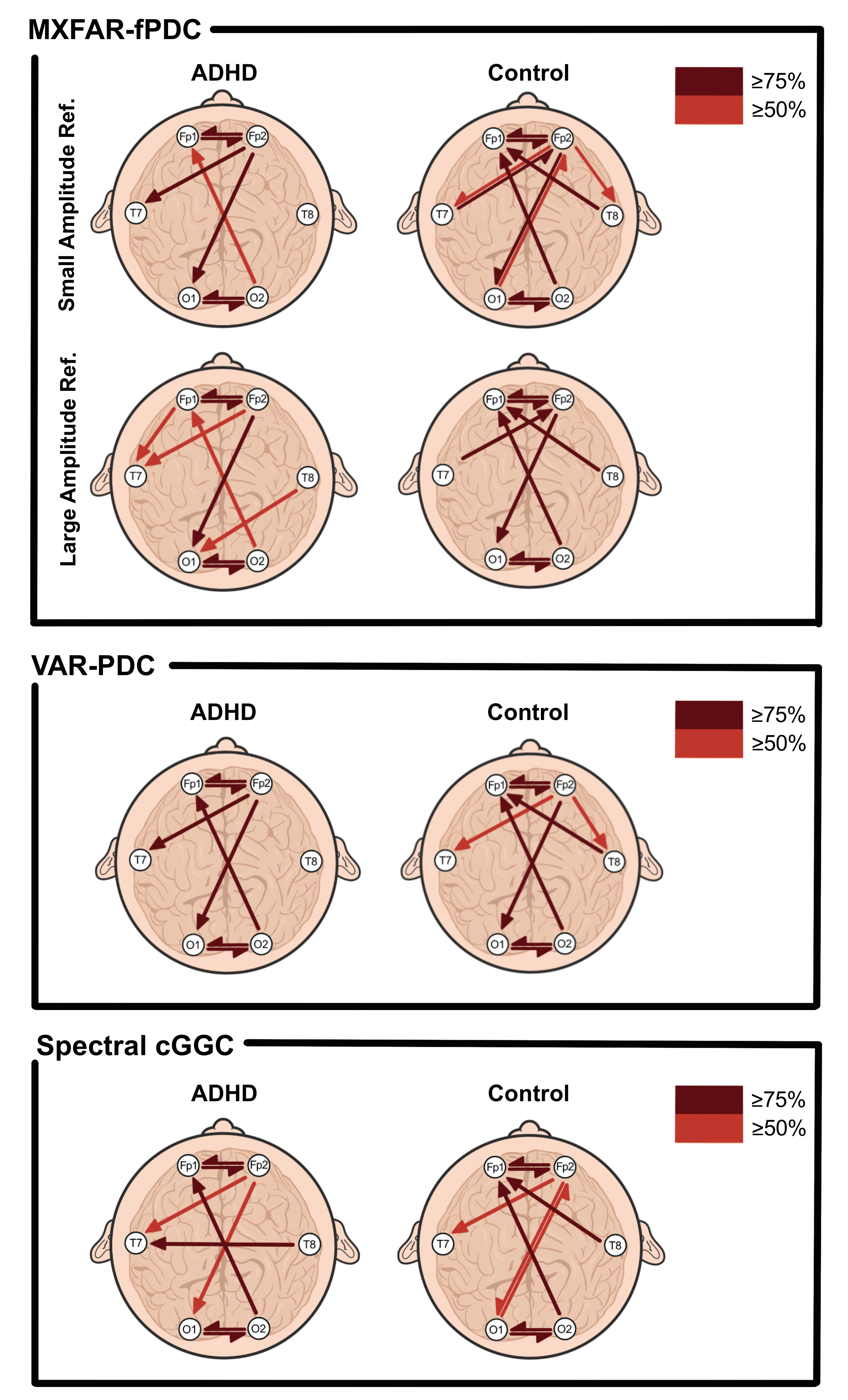}
	}
	\caption{Estimated EEG network \textbf{at low frequencies} (consisting of the delta, theta and alpha frequency bands) of the ADHD group (left) and the control group (right) based on three frameworks; MXFAR-\textit{f}PDC (first row panel), VAR-PDC (second row panel), and spectral cGGC (third row panel). Brain networks, based on the MXFAR-\textit{f}PDC frameworks, are derived given a small (top) or a large (bottom) amplitude of the reference signal (Cz at lag 6). Darker lines represent higher proportion of time windows where the directional link between two channels is prominent.}
	\label{fig:EEGSumm_DTA}
\end{figure}

\begin{figure}
	\centerline{
		\includegraphics[width=0.8\textwidth]{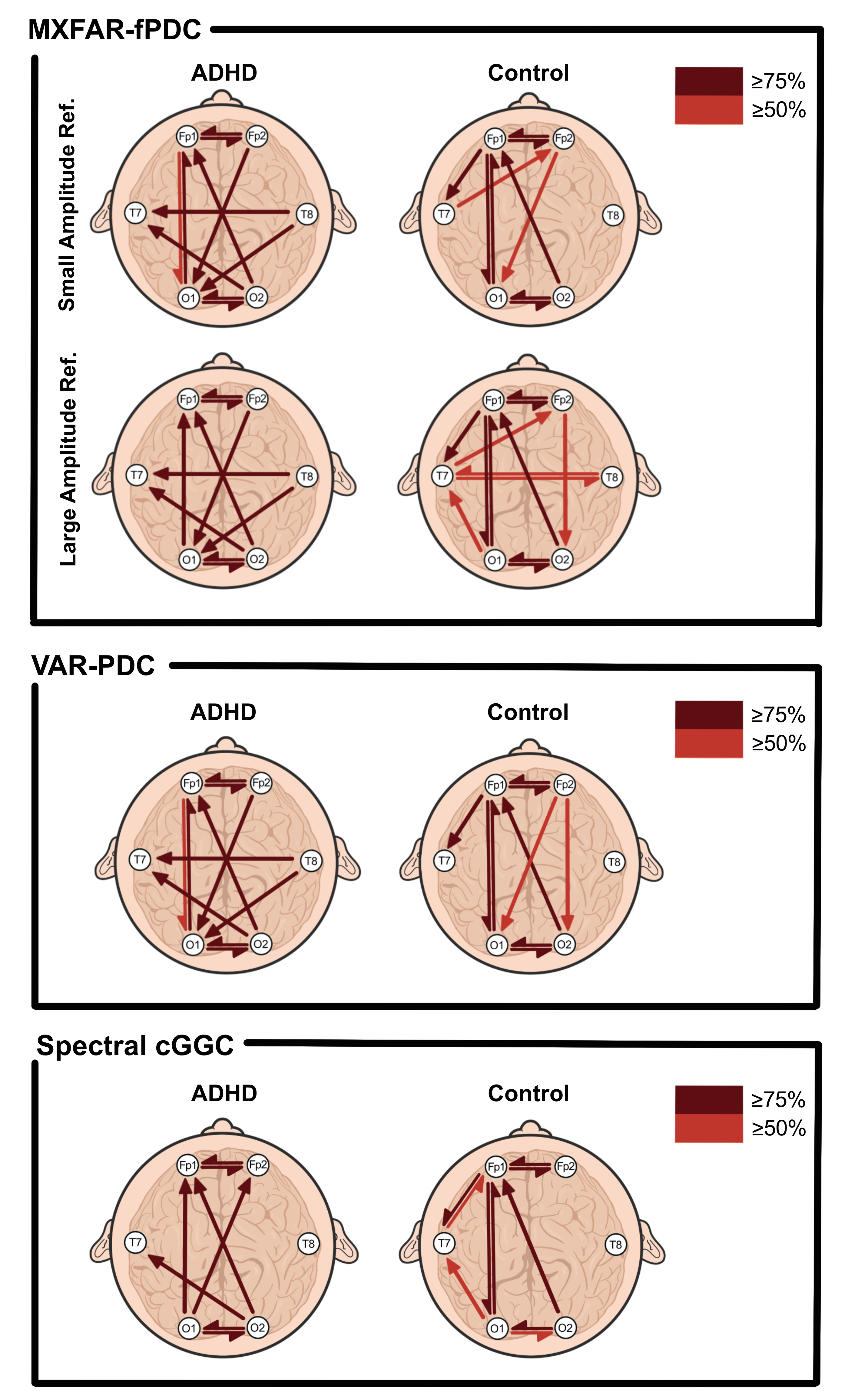}
	}
	\caption{Estimated EEG network \textbf{at high frequencies} (consisting of the beta, and gamma frequency bands) of the ADHD group (left) and the control group (right) based on three frameworks; MXFAR-\textit{f}PDC (first row panel), VAR-PDC (second row panel), and spectral cGGC (third row panel). Brain networks, based on the MXFAR-\textit{f}PDC frameworks, are derived given a small (top) or a large (bottom) amplitude of the reference signal (Cz at lag 6). Darker lines represent higher proportion of time windows where the directional link between two channels is prominent.}
	\label{fig:EEGSumm_BG}
\end{figure}

In each region, which consists of a finite grid of values for $\omega$ and $u$, we define the directional link $g \rightarrow j$ to be prominent (at the regional level) if at least ten percent of the individual \textit{f}PDC values in the region are prominent. This is to avoid spurious cases where a region would be deemed to be prominent only due to having only one or very few prominent individual \textit{f}PDC values. Lastly, to facilitate better interpretations for the derived connectivity network, we focus on prominent directional links $g \rightarrow j$ in the region that are \textit{consistent} throughout the duration of the visual experiment, i.e., connections that are prominent in at least 50 percent of the considered time windows. As a result, we argue that the derived brain connectivity networks based on these prominent and consistent directional links, summarized in Figures~\ref{fig:EEGSumm_DTA} and \ref{fig:EEGSumm_BG}, reflect the most critical connections associated with the visual tasks.

In comparison with our MXFAR-\textit{f}PDC framework, we perform analysis of the same EEG data based on two commonly-used methodologies for brain connectivity in the frequency domain \citep{cekic2018time}, namely, the (linear) PDC measure of \cite{baccala2001partial} and the spectral conditional Geweke-Granger causality (cGGC) measure of \cite{geweke1984measures}. The PDC measure, defined from a stationary linear VAR($p$) process, is the linear counterpart of our \textit{f}PDC metric while the spectral cGGC measure is the frequency decomposition of the conditional Granger causality measure derived from fitting an unrestricted and a restricted linear VAR($p$) models; for the formulation and estimation of the spectral cGGC metric, see the original work of \cite{geweke1984measures} or \cite{chen2006frequency}. Since PDC and spectral cGGC are both frequency-specific measures, results from their respective analysis are directly comparable with our \textit{f}PDC measure. However, due to the intrinsic linear formulation of PDC and spectral cGGC, we expect both frameworks to be inadequate in the presence of non-linearity in the dependence structure which, on the other hand, is properly addressed by the \textit{f}PDC framework.

Since we are dealing with multiple subjects, some aggregation techniques are applied before implementing the two linear frameworks. For the PDC, we summarize the linear coefficient matrices of the estimated individual VAR models by taking the average across all subjects with ADHD and all healthy controls. Then, the average linear coefficients are transformed into PDC values for the two groups of subjects. Similarly, for the spectral cGGC, the linear metric is calculated for each subject and is aggregated, by simple averaging, into the mean representatives for the group of subjects with ADHD and for the group of healthy controls. We focus on two regions of frequencies, which are the delta-theta-alpha band region and the beta-gamma band region, and apply similar strategies to define prominent and consistent directional links. That is, (i) we use the 0.8-quantile approach to determine prominent connections at each frequency; (ii) we consider the link to be prominent at the regional level if at least ten percent of the linear frequency-specific metrics are prominent in the region; and (iii) we concentrate on consistent directional links which are described as prominent links at the regional level in at least 50 percent of the considered time windows. By employing these strategies, we ensure that the three methodologies yield comparable brain networks that have connections of similar level of ``sparsity''. The derived brain connectivity network based on the PDC and spectral cGGC framework are also illustrated in Figures~\ref{fig:EEGSumm_DTA} and \ref{fig:EEGSumm_BG}.

There is a caveat to the brain connectivity networks extracted from prominent and consistent connections. Specifically, defining a directional link to be prominent does not necessarily imply its statistical significance. Thus, another interest is to derive brain networks based on \textit{significant} and consistent connections instead. However, testing for significance of the \textit{f}PDC estimates requires further theoretical investigation. One possible solution to this problem is to extend the frequency-specific threshold, developed by \cite{schelter2006testing}, for the PDC measure. Assuming similar conditions hold true, we outline a congruent strategy to define a non-linear frequency-specific threshold for our \textit{f}PDC measure and report the brain networks derived from significant and consistent connections in the Supplementary Material \citep{redondo2025suppmat}. Nevertheless, our discussions focus on prominent and consistent directional links which provide many interesting insights and novel findings about the difference in brain connectivity between subjects with and without ADHD.

\subsection{Results: Low Frequency Oscillations and Attention}

To our knowledge, our work is the first to develop the FAR framework for EEG data. Thus, the idea of having a reference signal that drives the dependence structure is new, which is one of our novel contributions. Moreover, our intention for implementing the MXFAR model is for exploratory purposes and hence, the following interpretations are the authors' speculations and still require verification from neuroscientists.

In Figure~\ref{fig:EEGSumm_DTA}, congruent connectivity patterns at low frequencies are extracted by the MXFAR-\textit{f}PDC framework for the ADHD group and for the healthy controls. For instance, regardless of the amplitude level of the reference signal, channels that have similar cognitive functions exhibit consistent ``two-way feedback" mechanism. Specifically, the pre-frontal channels Fp1 and Fp2, which are associated with problem solving, are connected in both directions (i.e., Fp1~$\longleftrightarrow$~Fp2) as well as the occipital channels O1 and O2 (i.e., O1~$\longleftrightarrow$~O2), which are linked with visual processing. Another similarity between the two groups is the complete loop of consistent information flow in the pre-frontal channels and the occipital channels (Fp1~$\rightarrow$~Fp2~$\rightarrow$~O1~$\rightarrow$~O2~$\rightarrow$~Fp1). We hypothesize that these connections are the ``minimum'' requirement for a subject to sustain attention and maintain focus while performing the visual task.

In addition to these similarities, our MXFAR-\textit{f}PDC framework reveals distinguishing features in connectivity at low frequencies between subjects with ADHD and healthy controls. In particular, given either small- or large-amplitude reference signals, the link O2~$\rightarrow$~Fp1 is less consistent in the ADHD group than the control group. This creates more frequent discontinuities in the pre-frontal-occipital cycle for subjects with ADHD which may have negative impacts in their capacity to maintain focus. Furthermore, in the ADHD group, the temporal channels act as receivers of information from the pre-frontal channels. For example, when past signals from the reference channel Cz exhibit small amplitudes, Fp2~$\rightarrow$~T7 consistently occurs, while connections Fp1~$\rightarrow$~T7 and Fp2~$\rightarrow$~T7 appear given large-amplitude reference signals. Another distinct connectivity feature at low frequencies among subjects with ADHD is the asymmetry between the left and right hemispheres of their brain network. Especially when large-amplitude signals from Cz are observed, the degree of asymmetry increases with the presence of additional connections Fp1~$\rightarrow$~T7 and T8~$\rightarrow$~O1, hence suggesting non-linear dependence structure at low frequencies for subjects with ADHD. Such features may be linked with instabilities in brain connectivity that translate to shorter attention span among subjects in the ADHD group.

By contrast, for the healthy controls, the temporal channels consistently transfer information to, rather than receive information from, the pre-frontal channels. Specifically, at any amplitude level of the reference signal, the directional links T7~$\rightarrow$~Fp2 and T8~$\rightarrow$~Fp1 remain prominent and consistent in the duration of the visual experiment. Additionally, when there is less activity in other parts of the cortex as reflected by small-amplitude signals from Cz, more notable interactions between the temporal and pre-frontal channels occur such as the two-way feedback Fp2~$\longleftrightarrow$~T7 and the complete pre-frontal-temporal loop Fp1~$\rightarrow$~Fp2~$\rightarrow$~T8~$\rightarrow$~Fp1. This general role of the temporal channels for subjects in the control group, as transmitters of information to the pre-frontal channels, may be a critical feature for maintaining focus and sustaining attention.

Another difference is that, at low frequencies, unlike the asymmetric network derived from the ADHD group, the control group exhibits more symmetric connectivity patterns with respect to the brain hemispheres. That is, we see congruent directional links originating from the left (Fp1, O1, and T7) and right (Fp2, O2, and T8) channels. Moreover, such symmetric network remains consistent regardless of the amplitude level of the reference signal, thus suggesting a stable approximately-linear dependence structure in the brain network of healthy subjects. We speculate that having symmetry and stability in connectivity helps healthy subjects to stay engaged and focused in completing the visual experiment.

In comparison, common directional links are revealed by the PDC and spectral cGGC frameworks, including the two-way feedback between the pre-frontal channels (Fp1~$\longleftrightarrow$~Fp2) and between the occipital channels (O1~$\longleftrightarrow$~O2), and the pre-frontal-occipital information loop (Fp1~$\rightarrow$~Fp2~$\rightarrow$~O1~$\rightarrow$~O2~$\rightarrow$~Fp1). This is congruent with the results from our non-linear framework which provides additional support on the hypothesis of having a required minimum set of connections to sustain attention. However, the two linear frameworks do not fully capture the differences between the two groups. In fact, the only differences between the ADHD group and the control group, extracted by the PDC metric, are the connections between the pre-frontal channels and right temporal channel (i.e., T8~$\rightarrow$~Fp1 and Fp2~$\rightarrow$~T8) among the healthy subjects. On the other hand, results from the spectral cGGC framework share similar interpretations as the linear PDC analysis; however, they differ slightly on the consistency of the connections in the derived brain networks. Hence, whenever non-linearity exists in the dependence structure and is driven by a reference signal, our proposed framework holds an advantage over linear frameworks such as the PDC and the spectral cGGC, since it captures complex cross-channel interactions more realistically.

\subsection{Results: High Frequency Oscillations and Information Processing}

Figure~\ref{fig:EEGSumm_BG} illustrates the prominent and consistent connections, at high frequencies, in the brain network of subjects with ADHD and healthy controls. As revealed by the MXFAR-\textit{f}PDC framework for both groups, channels with similar cognitive functions exhibit two-way interactions, e.g., the pre-frontal channels (Fp1~$\longleftrightarrow$~Fp2) and the occipital channels (O1~$\longleftrightarrow$~O2). Given that these connections also appear at low frequencies, we stress that the pre-frontal and occipital channels have an essential involvement during execution of visual-cognitive tasks. Additionally, consistent information transfer from the occipital channels to the pre-frontal channels (i.e., O1~$\rightarrow$~Fp1 and O2~$\rightarrow$~Fp1) are extracted from both the ADHD group and the control group, which occur at any amplitude level of the reference signal. We expect these directional links to be prominent and consistent as counting characters from a flashed image requires processing visual inputs (from the occipital channels) and invoking concentration-related or problem solving skills (from the pre-frontal channels). Since all subjects, with or without ADHD, perform the same experiment, such similarity in their connectivity structure points to the presence of a set of ``baseline'' pathways that carry out information processing for visual-related functions.

However, the brain connectivity networks derived from the ADHD group and from the control group differ with respect to the impact of the reference signal at high frequencies. In fact, the former approximate a linear dependence structure which is unaffected by the amplitude level of the reference signal, while the latter exhibits non-linear connectivity patterns driven by the reference signal. Furthermore, another difference between the two brain networks is the flow of information from the pre-frontal to the occipital channels. Particularly, the link Fp2~$\rightarrow$~O1 consistently appears for subjects with ADHD, while healthy subjects exhibit consistent Fp1~$\rightarrow$~O1 connection. Although the two links are different, both allow for a feedback mechanism between the pre-frontal and the occipital channels, e.g., Fp1~$\rightarrow$~Fp2~$\rightarrow$~O1~$\rightarrow$~Fp1 for the ADHD group, and Fp1~$\longleftrightarrow$~O1 for the control group. However, we see that healthy subjects have a more direct information transfer pathway between the pre-frontal and the occipital channel than subjects with ADHD. 

An additional dissimilarity between the two groups, highlighted by our MXFAR-\textit{f}PDC framework, is the participation of the temporal channels at high frequencies. For ADHD subjects, there is transfer of information from the right temporal to the left temporal channel that passes through the occipital channels (i.e., T8~$\rightarrow$~O1~$\rightarrow$~O2~$\rightarrow$~T7), which remains consistent regardless of how large or small past activities are in other parts of the cortex (as reflected by the signals from channel Cz). We postulate that this information pathway is associated with the phenomenon where subjects with ADHD consciously remind themselves to continue with the visual task whenever their attention starts to deviate away from the experiment. Similarly, the temporal channels are more involved with the pre-frontal channels for the healthy controls at high frequencies. For instance, at any amplitude level of the reference signal, a consistent information cycle occurs between the pre-frontal channels and the left temporal channel (Fp1~$\rightarrow$~T7~$\rightarrow$~Fp2~$\rightarrow$~Fp1). Moreover, when past signals from the reference channel Cz exhibit large amplitudes, several indirect pathways from the occipital to the pre-frontal channels, through the left temporal channel, which includes the connection O1~$\rightarrow$~T7~$\rightarrow$~Fp2, becomes prominent and consistent. We postulate that information from visual inputs (occipital channels) passing through the temporal channels, before reaching the problem solving units (pre-frontal channels), resembles the utilization of stored memory. That is, whenever a subject remembers seeing a similar image, the task of counting characters becomes easier. This leads to the efficiency of healthy subjects in performing the visual experiment, with an average completion time of 121.53 seconds and standard deviation of 29.36 seconds, as compared to the subjects with ADHD, with an average completion time and standard deviation of 149.35 seconds and 55.22 seconds, respectively.

The linear PDC framework demonstrates a similar brain connectivity network at high frequencies for the ADHD group. This provides further evidence on the linearity in cross-channel dependence among subjects with ADHD, as well as the capability of our MXFAR-\textit{f}PDC framework to extract, not only non-linear, but also linear connectivity patterns. Moreover, similar connections at high frequencies are highlighted by the linear PDC and the spectral cGGC frameworks including the two-way feedback between the pre-frontal channels and between the occipital channels, and the information transfer from the occipital to the pre-frontal channels (O1~$\rightarrow$~Fp1 and O2~$\rightarrow$~Fp1). However, the derived brain networks at high frequencies using the spectral cGGC framework have several missing directional links and connections with less consistency. Despite being frequency-specific, the spectral cGGC measure is derived from the definition of Granger causality which is based on variances of future value predictions. In the analysis of EEG signals where their actual values do not have a direct meaning, prediction of future values may not be useful, thus explaining why the spectral cGGC framework is unable to extract some of the prominent and consistent directional links. Furthermore, the drawback of the linear PDC and the spectral cGGC frameworks is that they fail to capture the non-linear dependence structure among the healthy subjects which may explain the efficiency of healthy individuals in completing the visual-cognitive experiment. Hence, our MXFAR-\textit{f}PDC framework is a more appropriate tool that addresses the limitations of existing linear methodologies in the presence of a reference signal that drives the non-linear dependence structure.

\section{Conclusion}\label{chap:conclusion}

\noindent One of the advantages of the MXFAR model is its flexibility: functional coefficients can take either additive or multiplicative subject-specific effects. Furthermore, the proposed estimation approach does not require knowledge of the true functional form of the coefficients nor its governing parameters. Instead, the functions are estimated by assuming a locally linear structure with random effects. Thus, the inference method provides good estimates of the true mean functions and account for between-subject variation. Moreover, the model can also accommodate various group-specific features making our model very flexible, and hence useful for differentiating the brain connectivity networks of ADHD patients against the healthy controls.

Another contribution is the development of the \textit{f}PDC metric, a new non-linear dependence measure in the frequency domain. As a non-linear generalization of the original PDC, the \textit{f}PDC measure quantifies the direction and magnitude of information flow between nodes in a network as a joint function of frequency oscillations and fluctuations of a reference signal. In addition, by taking advantage of the MXFAR model, an aggregated dynamic dependence structure may be derived from translating the mean functional coefficients into the \textit{f}PDC measure, which naturally incorporates subject-specific variability. Thus, compared to existing methodologies based on linear spectral metrics, the proposed non-linear frequency-specific measure enables for capturing more sensible connectivity features when analyzing complex dynamic systems such as brain functional networks.

The MXFAR-\textit{f}PDC framework produced novel and interesting findings in the analysis of the ADHD EEG data. Using the recordings from the central channel Cz as a reference signal that represents the general activity in the remaining parts of the neural cortex, the proposed framework was able to identify more pronounced linear and non-linear dependence structures for the ADHD group and the healthy control group. At low frequencies, healthy controls demonstrate symmetric and approximately linear connectivity patterns while subjects with ADHD yield an asymmetric dependence structure especially when the reference signals exhibit large amplitudes. The stability in connectivity may be associated with the ability to maintain focus whereas the imbalance in the network may be a reflection of having shorter attention span. By contrast, cross-channel dependence at high frequencies reveal the opposite conclusion, i.e., the derived brain network for the ADHD group indicates a linear structure while healthy subjects display connectivity features that are driven by the amplitude level of the reference signal. This non-linear brain network among healthy controls may be related to the role of the temporal channels when utilizing stored memory. Further investigation on the interaction of the temporal channels with other brain regions during cognition may lead to identification of possible biomarkers that differentiate subjects with ADHD from healthy individuals.

Finally, even though the proposed mixed-effects non-linear model is here applied to EEG signals, the modeling framework is very general and may have potential applications in entirely different contexts. For example, Gross Domestic Product (GDP) from different countries may be explained through the MXFAR model, where GDPs from top producing countries may serve as the reference signals as they drive the global economy. The same model may be useful for stock market prices where leading companies dictate the direction of the market. Practitioners may employ our MXFAR methodology using the \texttt{mxfar} package, downloadable through the Github repository \url{https://github.com/ptredondo/mxfar}.






\begin{acks}[Acknowledgments]
The authors thank Sarah Bernadette Aracid for the elaborate artworks (Figures~\ref{fig:eegsamp}, \ref{fig:EEGSumm_DTA} and \ref{fig:EEGSumm_BG}).
\end{acks}



\begin{supplement}
\stitle{Supplementary Material for ``Functional-Coefficient Models for Multivariate Time Series in Designed Experiments: with Applications to Brain Signals"}
\sdescription{Includes a concise summary of the univariate and multivariate FAR models, the definition of a univariate MXFAR process, the results of the extensive numerical experiment, the results from the sensitivity analysis for the choice of channels in the data application, the outline for defining a non-linear (point-wise) frequency-specific significance threshold for the functional partial directed coherence estimates, and additional results and plots for the numerical experiments and the analysis of EEG data.}
\end{supplement}

\begin{supplement}
\stitle{\texttt{R} Codes and EEG Data for ``Functional-Coefficient Models for Multivariate Time Series in Designed Experiments: with Applications to Brain Signals"}
\sdescription{Includes the data and codes to reproduce key results of the EEG data analysis, which are accessible in the Github repository \url{https://github.com/ptredondo/mxfar}.}
\end{supplement}


\bibliographystyle{imsart-nameyear} 
\bibliography{bibliography}       

\end{document}